\documentclass[12pt]{article}
\usepackage{amsmath}
\usepackage{amssymb} 
\usepackage{amsthm}
\usepackage{bbm}
\usepackage{graphicx,epsf}
\usepackage{epstopdf}
\usepackage{enumerate}
\usepackage{caption}
\usepackage{natbib} 
\usepackage{bm}
\newcommand{\blind}{0}

\newcommand{\med}{\mathop{\mbox{med}}}

\newcommand{\DO}{\mbox{DO}}
\newcommand{\MAD}{\mbox{MAD}}
\newcommand{\LDO}{\mbox{LDO}}

\newcommand{\soa}{\mbox{s}_{o,a}}
\newcommand{\sam}{\mbox{s}_a}
\newcommand{\sob}{\mbox{s}_{o,b}}
\newcommand{\sbm}{\mbox{s}_b}
\newcommand{\R}{\mathbb{R}}
\newcommand{\one}{\mathbbm{1}}
\newcommand{\balpha}{\boldsymbol \alpha}
\newcommand{\bv}{\boldsymbol v}
\newcommand{\bx}{\boldsymbol x}
\newcommand{\by}{\boldsymbol y}
\newcommand{\bY}{\boldsymbol Y}

\newtheorem{proposition}{Proposition}
\newtheorem{lemma}{Lemma}
\DeclareMathOperator{\sign}{sign}

\begin{document}

\def\spacingset#1{\renewcommand{\baselinestretch}%
{#1}\small\normalsize} \spacingset{1}

%%%%%%%%%%%%%%%%%%%%%%%%%%%%%%%%%%%%%%%%%%%%%%%%%%%%%%%%%%%

\if0\blind
{
  \title{\bf A Measure of Directional Outlyingness\\
	           with Applications to Image Data and Video}		
  \author{Peter J. Rousseeuw, Jakob Raymaekers and 
	        Mia Hubert\thanks{
	  This research	has been supported by 
		projects of Internal Funds KU Leuven.
    The authors are grateful for interesting discussions 
		with Pieter Segaert.} \hspace{.1cm} \\
		Department of Mathematics, KU Leuven, Belgium}
	\date{March 3, 2017}
  \maketitle
} \fi

\if1\blind
{
  \bigskip
  \bigskip
  \bigskip
  \begin{center}
	  {\LARGE\bf A Measure of Directional Outlyingness\\
	             with Applications to Image Data and Video}			
\end{center}
  \medskip
} \fi

\bigskip
\begin{abstract}
Functional data covers a wide range of data types.
They all have in common that the observed objects are 
functions of of a univariate argument (e.g. time or
wavelength) or a multivariate argument (say, a spatial 
position).
These functions take on values which can in turn be 
univariate (such as the absorbance level) or 
multivariate (such as the red/green/blue color levels of 
an image).
In practice it is important to be able to detect outliers
in such data.
For this purpose we introduce a new measure of outlyingness 
that we compute at each gridpoint of the functions' domain. 
The proposed {\it directional outlyingness} (DO) measure
accounts for skewness in the data and only requires 
$\mathcal{O}(n)$ computation time per direction. 
We derive the influence function of the DO and compute a cutoff for outlier detection. 
The resulting heatmap and functional outlier map 
reflect local and global outlyingness of a function.
To illustrate the performance of the method on real data it 
is applied to spectra, MRI images, and video surveillance data.
\end{abstract}

\vskip0.3cm
\noindent%
\vfill

\newpage
\spacingset{1.45}
\section{Introduction}
\label{sec:intro}
Functional data 
analysis~\citep{Ramsay:BookFDA,Ferraty:BookFDA}
is a rapidly growing research area.
Often the focus is on functions with a univariate domain, 
such as time series or spectra. 
The function values may be multivariate, such as 
temperatures measured at 3, 9 and 12 cm below ground
\citep{Berrendero:PCAFunc} or human ECG data measured
at 8 different places on the body \citep{Pigoli:Wavelets}.
In this paper we will also consider functions whose 
{\it domain} is multivariate.
In particular, images and surfaces are functions on a 
bivariate domain.
Our methods generalize to higher-dimensional domains, e.g.
the voxels of a 3D-image of a human brain are defined on
a trivariate domain.

Detecting outliers in functional data is an important task.
Recent developments include the approaches of
\cite{Febrero:OutlFunc} and \cite{Hyndman:Bagplot}.
\cite{Sun:FuncBoxplots} proposed the functional boxplot, and
\cite{Arribas:Outliergram} developed the outliergram.
Our approach is somewhat different.
To detect outlying functions or outlying parts of a
function (in a data set consisting of several functions)
we will look at its (possibly multivariate) function value
in every time point/pixel/voxel/... of its domain.
For this purpose we need a tool that assigns a measure
of outlyingness to every data point in a multivariate
non-functional sample.
A popular measure is the Stahel-Donoho outlyingness (SDO)
due to \cite{Stahel:SDEst} and \cite{Donoho:SDest} which 
works best when the distribution of the inliers is roughly
elliptical.
However, it is less suited for skewed data.
To address this issue, \cite{Brys:RobICA}
proposed the (Skewness-) Adjusted Outlyingness (AO)
which takes the skewness of the underlying distribution
into account.
However, the AO has two drawbacks.
The first is that the AO scale has a large bias
as soon as the contamination fraction exceeds 10\%. 
Furthermore, its computation time is $\mathcal{O}(n\log(n))$ 
per direction due to its rather involved construction.

To remedy these deficiencies we propose a new measure in
this paper, the {\it Directional Outlyingness} (DO). 
The DO also takes the skewness of the underlying distribution
into account, by the intuitive idea of splitting a 
univariate dataset in two half samples around the median.
The AO incorporates a more robust scale estimator, which
requires only $\mathcal{O}(n)$ operations. 

Section 2 defines the DO, investigates its
theoretical properties and illustrates it on univariate,
bivariate and spectral data. 
Section 3 derives a cutoff value for the DO and applies
it to outlier detection. 
It also extends the functional outlier map of
\cite{Hubert:MFOD} to the DO, and in it constructs
a curve separating outliers from inliers.
Section 4 shows an application to MRI images,
and Section 5 analyzes video data.
Section 6 contains simulations in various settings, to
study the behavior of DO and compare its performance to
other methods. 
Section 7 concludes.

\section{A Notion of Directional Outlyingness}
\label{sec:meth}
\subsection{Univariate Setting}
In the univariate setting, the Stahel-Donoho outlyingness of 
a point $y$ relative to a sample $Y=\{y_1,\ldots,y_n\}$ is 
defined as
\begin{equation}\label{eq:SDO_univ}
  \text{SDO}(y;Y)= \frac{|\; y - \med(Y) \;|}{\MAD(Y)}
\end{equation}
\noindent
where the denominator is the median absolute deviation (MAD) 
of the sample, given by
$\MAD(Y)=\med_i(|\; y_i-\med_j(y_j)|)/\Phi^{-1}(0.75)$
where $\Phi$ is the standard normal cdf.
The SDO is affine invariant, meaning that it remains the same
when a constant is added to $Y$ and $y$, and also when they 
are multiplied by a nonzero constant.

A limitation of the SDO is that it implicitly assumes the 
inliers (i.e. the non-outliers) to be roughly symmetrically
distributed. But when the inliers have a skewed distribution,
using the MAD as a single measure of scale does not capture 
the asymmetry. 
For instance, when the data stem from a right-skewed
distribution, the SDO may become large for inliers on the
right hand side, and not large enough for actual outliers
on the left hand side.

This observation led to the (skewness-) adjusted outlyingness 
(AO) proposed by~\cite{Brys:RobICA}.
This notion employs a robust measure of skewness called the
medcouple~\citep{Brys:RobSkew}, which however requires
$\mathcal{O}(n \log(n))$ computation time. 
Moreover, we will see in the next subsection that it leads
to a rather large explosion bias.

In this paper we propose the notion of 
{\it directional outlyingness} (DO) which also takes the
potential skewness of the underlying distribution into
account, while attaining a smaller computation time and bias.
The main idea is to split the sample into two half samples, 
and then to apply a robust scale estimator to each of them. 

More precisely, 
let $y_1 \leqslant y_2 \leqslant \ldots \leqslant y_n$ be 
a univariate sample.
(The actual algorithm does not require sorting the data.)
We then construct two subsamples of size
$h=\left\lfloor \frac{n+1}{2}\right\rfloor$ as follows.
For even $n$ we take $Y_{a}=\{y_{h+1},\ldots,y_n\}$ and 
$Y_{b}=\{y_1,\ldots,y_h\}$ where the subscripts $a$ and $b$
stand for {\bf a}bove and {\bf b}elow the median.
For odd $n$ we put $Y_{a}=\{y_{h},\ldots,y_n\}$ and 
$Y_{b}=\{y_1,\ldots,y_h\}$ so that $Y_a$ and $Y_b$ share
one data point and have the same size.

Next, we compute a scale estimate for each subsample.
Among many available robust estimators we choose a one-step
M-estimator with Huber $\rho$-function due to its fast
computation and favorable properties.
We first compute initial scale estimates
\begin{equation*}
  \soa(Y)=\med(Z_a)/\Phi^{-1}(0.75)
	\hskip0.5cm \mbox{and} \hskip0.5cm
  \sob(Y)=\med(Z_b)/\Phi^{-1}(0.75) 
\end{equation*}
where $Z_a = Y_a - \med(Y)$ and $Z_b = \med(Y) - Y_b$\;
and where $\Phi^{-1}(0.75)$ ensures consistency for gaussian 
data.
The one-step M-estimates are then given by
\begin{equation}\label{eq:sasb}
	\begin{array}{ll}
\sam(Y)=&\soa(Y)\sqrt{\; \frac{1}{2\alpha h}\; \displaystyle
 \sum_{z_i \in Z_a}{\rho_c \left(\frac{z_i}{\soa(Y)}\right)}}\\\\
\sbm(Y)=&\sob(Y)\sqrt{\; \frac{1}{2\alpha h}\; \displaystyle
 \sum_{z_i \in Z_b}{\rho_c \left(\frac{z_i}{\sob(Y)}\right)}}\\
	\end{array} 
\end{equation}
where again $h=\left\lfloor \frac{n+1}{2}\right\rfloor$ and where
$\alpha = \int_{0}^{\infty}{\rho_c(x)d\Phi(x)}$.
Here $\rho_c$ denotes the Huber rho function for scale 
$\rho_c(t)=\left(\frac{t}{c}\right)^2 \one_{[-c,c]}
 +\one_{(-\infty,c) \cup (c,\infty)}$\; 
with $c$ a tuning parameter regulating the trade-off between 
efficiency and bias.

Finally, the directional outlyingness (DO) of a point $y$ 
relative to a univariate sample 
$Y=\{y_1,\ldots,y_n\}$ is given by
\begin{equation}\label{eq:DO_univ}
\DO(y;Y)=
  \left\{
	\begin{array}{ll}
		 \frac{y-\med(Y)}{\sam(Y)}
		      & \;\; \mbox{ if } y \geqslant \med(Y) \\\\
			\frac{\med(Y) - y}{\sbm(Y)}
		      & \;\; \mbox{ if } y \leqslant \med(Y) \;\;.\\
	\end{array} 
  \right. 
\end{equation}
Note that DO is affine invariant. In particular, flipping 
the signs of $Y$ and $y$ interchanges $\sam$ and $\sbm$
which results in $\DO(-y;-Y)=\DO(y;Y)$.

\begin{figure}[!htb]
\centering
\includegraphics[width=0.6\textwidth]
                {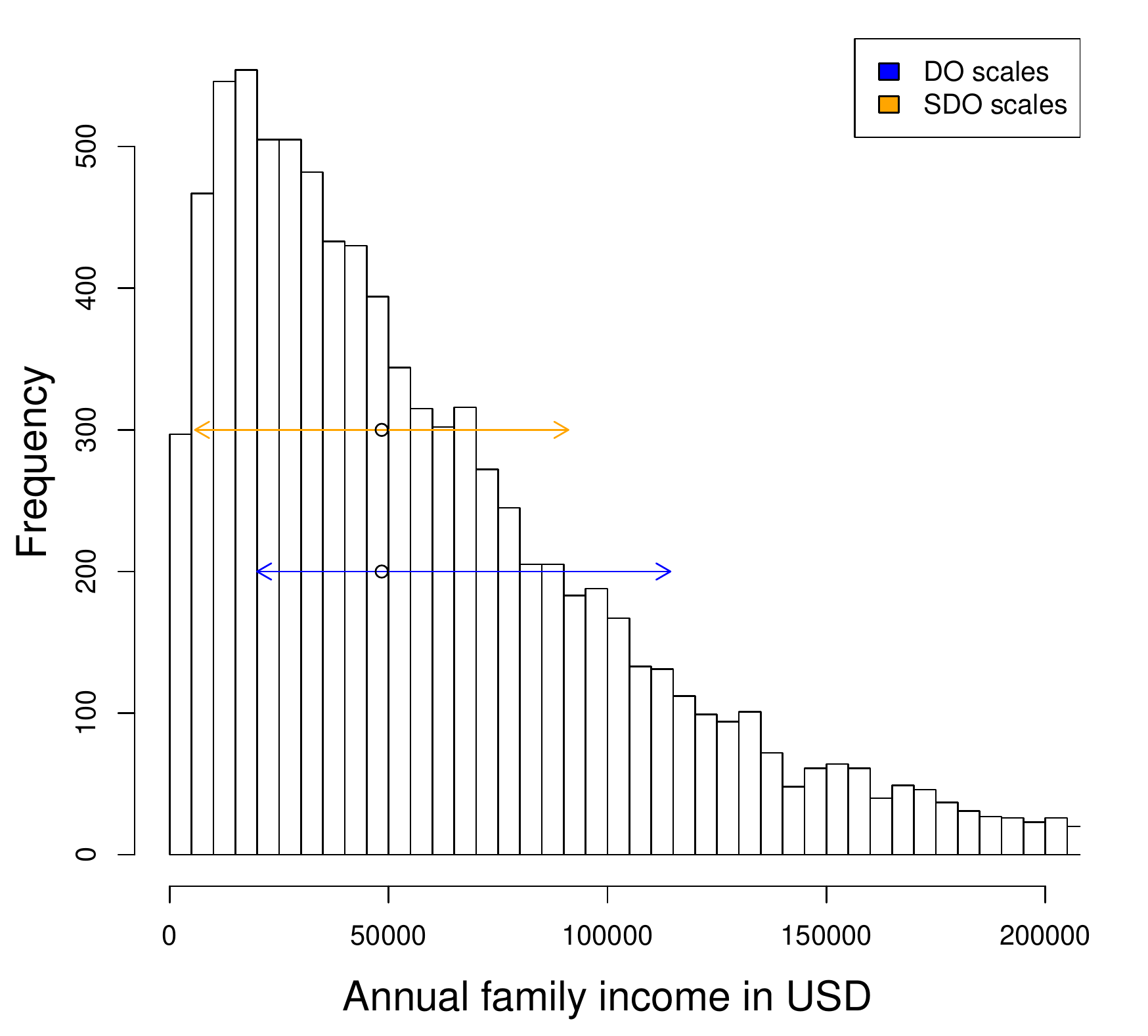}
\vskip-0.2cm
\caption{Scale estimates of the family income data.}
		\label{fig:scales}
\end{figure}

Figure~\ref{fig:scales} illustrates the denominators of 
the SDO and DO expressions on the family income dataset 
from \textsl{https://psidonline.isr.umich.edu}
which contains 8962 strictly positive incomes in the tax 
year 2012.
Their histogram is clearly right-skewed. 
The MAD in the denominator of SDO equals $\$42,650$ and 
is used both to the left and to the right of the median, 
as depicted by the orange arrows.
For the DO the `above' scale $\sam=\$58,681$ exceeds the
`below' scale $\sbm=\$35,737$ (blue arrows). 
Therefore, a point to the right of the median will have
a lower DO than a point to the left at the same distance
to the median.
This is a desirable property in view of the
difference between the left and right tails.

\subsection{Robustness Properties}
Let us now study the robustness properties of the scales
$\sam$ and $\sbm$ and the resulting DO.\newline
It will be convenient to write $\sam$ and $\sbm$ as 
functionals of the data distribution $F$:
\begin{equation} \label{eq:funcsasb}
\begin{array}{ll}
\sam^2(F)&=\frac{\soa^2(F)}{\alpha} \int_{\med(F)}^{\infty}\;
  {\rho_c\left(\frac{x-\med(F)}{\soa(F)}\right) dF(x)}\\
	\\
\sbm^2(F)&=\frac{\sob^2(F)}{\alpha} \int_{-\infty}^{\med(F)}\;
  {\rho_c\left(\frac{\med(F)-x}{\sob(F)}\right) dF(x)}
\end{array}
\end{equation}
where $\rho_c$ is the Huber $\rho$-function.

We will first focus on the worst-case bias of $\sam$ due
to a fraction $\varepsilon$ of contamination, 
following~\cite{Martin:Biasscale}.
At a given distribution $F$, the explosion bias curve 
of $\sam$ is defined as
\begin{equation*}
B^+(\varepsilon,\sam,F)=
     \sup_{G \in \mathcal{F}_{\varepsilon}}(\sam(G))
\end{equation*}
where $\mathcal{F}_{\varepsilon}=
       \{G:G=(1-\varepsilon)F+\varepsilon H\}$ 
in which $H$ can be any distribution.
The implosion bias curve is defined similarly as
\begin{equation*}
B^-(\varepsilon,\sam,F)=
     \inf_{G \in \mathcal{F}_{\varepsilon}}(\sam(G))\;\;.
\end{equation*}

From here onward we will assume that $F$ is symmetric 
about some center $m$ and has a continuous density $f(x)$ 
which is strictly decreasing in $x>m$.
In order to derive the explosion and implosion bias we 
require the following lemma (all proofs can be found in 
the supplementary material):
\begin{lemma}
(i) For fixed $\mu$ it holds that\;
 $t^2 \int_{\mu}^{\infty}
  {\rho_c\left(\frac{x-\mu}{t}\right) dF(x)}$ 
is strictly increasing in $t>0$;\\ 
(ii) For fixed $\sigma>0$ it holds that\;
	$\sigma^2 \int_{t}^{\infty}
	{\rho_c\left(\frac{x-t}{\sigma}\right) dF(x)}$ 
is strictly decreasing in $t$. 
\end{lemma}

\begin{proposition}
For any $0 < \varepsilon < 0.25$ the implosion bias of 
$\sam$ is given by
\begin{equation*}
\begin{array}{ll}
  B^-(\varepsilon,\sam,F)^2&=
	\frac{1}{\alpha} B^-(\varepsilon,\soa,F)^2
	\left\{ (1-\varepsilon) 
	\int\displaylimits_{B^+(\varepsilon,\med,F)}^{\infty}
	{\rho_c\left(\frac{x-B^+(\varepsilon,\med,F)}
	{B^-(\varepsilon,\soa,F)}\right)dF(x)} \right\}\\
\end{array}
\end{equation*}
where
\begin{equation*}
\begin{array}{lll}
  B^+(\varepsilon,\med,F)&=&F^{-1}\left(
	  \frac{1}{2(1-\varepsilon)}\right)\\
  B^-(\varepsilon,\soa,F)&=&\frac{1}{\Phi^{-1}\left(
	  \frac{3}{4}\right)}\left\{F^{-1}\left(
		\frac{3-4\varepsilon}{4(1-\varepsilon)}\right)
		-F^{-1}\left(
		\frac{1}{2(1-\varepsilon)}\right)\right\}\;.
\end{array}
\end{equation*}
\end{proposition}

In fact, the implosion bias of $\sam$ is reached 
when $H = \Delta \left(F^{-1}
          \left(\frac{1}{2(1-\varepsilon)}\right)\right)$
is the distribution that puts all its mass in 
the point $F^{-1}\left(\frac{1}{2(1-\varepsilon)}\right)$. 
Note that the {\it implosion breakdown value} of $\sam$ 
is 25\% because for $\varepsilon \rightarrow 0.25$ we 
obtain $\sam \rightarrow 0$.

\begin{proposition}
For any $0 < \varepsilon < 0.25$ the explosion bias of 
$\sam$ is given by
\begin{equation*}
\begin{array}{ll}
  B^+(\varepsilon,\sam,F)^2&=
    \frac{1}{\alpha} B^+(\varepsilon,\soa,F)^2
		\left\{(1-\varepsilon) 
		\int\displaylimits_{B^+(\varepsilon,\med,F)}^{\infty}
		{\rho_c\left(\frac{x-B^+(\varepsilon,\med,F)}
		{B^+(\varepsilon,\soa,F)}\right)dF(x)}+\varepsilon
		\right\}\\
\end{array}
\end{equation*}
where
\begin{equation*}
\begin{array}{lll}
  B^+(\varepsilon,\soa,F)&=&\frac{1}{\Phi^{-1}
  \left(\frac{3}{4}\right)}\left\{F^{-1}
  \left(\frac{3}{4(1-\varepsilon)}\right)-F^{-1}
	\left(\frac{1}{2(1-\varepsilon)}\right)\right\}\;.\\
\end{array}
\end{equation*}
\end{proposition}

The explosion bias of $\sam$ is reached at all distributions 
$F_{\varepsilon}=(1-\varepsilon)F+\varepsilon\Delta(d)$ for which
$d > B^{+}(\varepsilon,med,F) + cB^{+}(\varepsilon,s_{o,a},F)$
which ensures that $d$ lands on the constant part of $\rho_c$\;.
For $\varepsilon \rightarrow 0.25$ we find $d \rightarrow \infty$
and $\sam \rightarrow \infty$, so the
{\it explosion breakdown value} of $\sam$ is 25\%\;. 

\begin{figure}[!htb]
\centering
\includegraphics[width=0.6\textwidth]
                {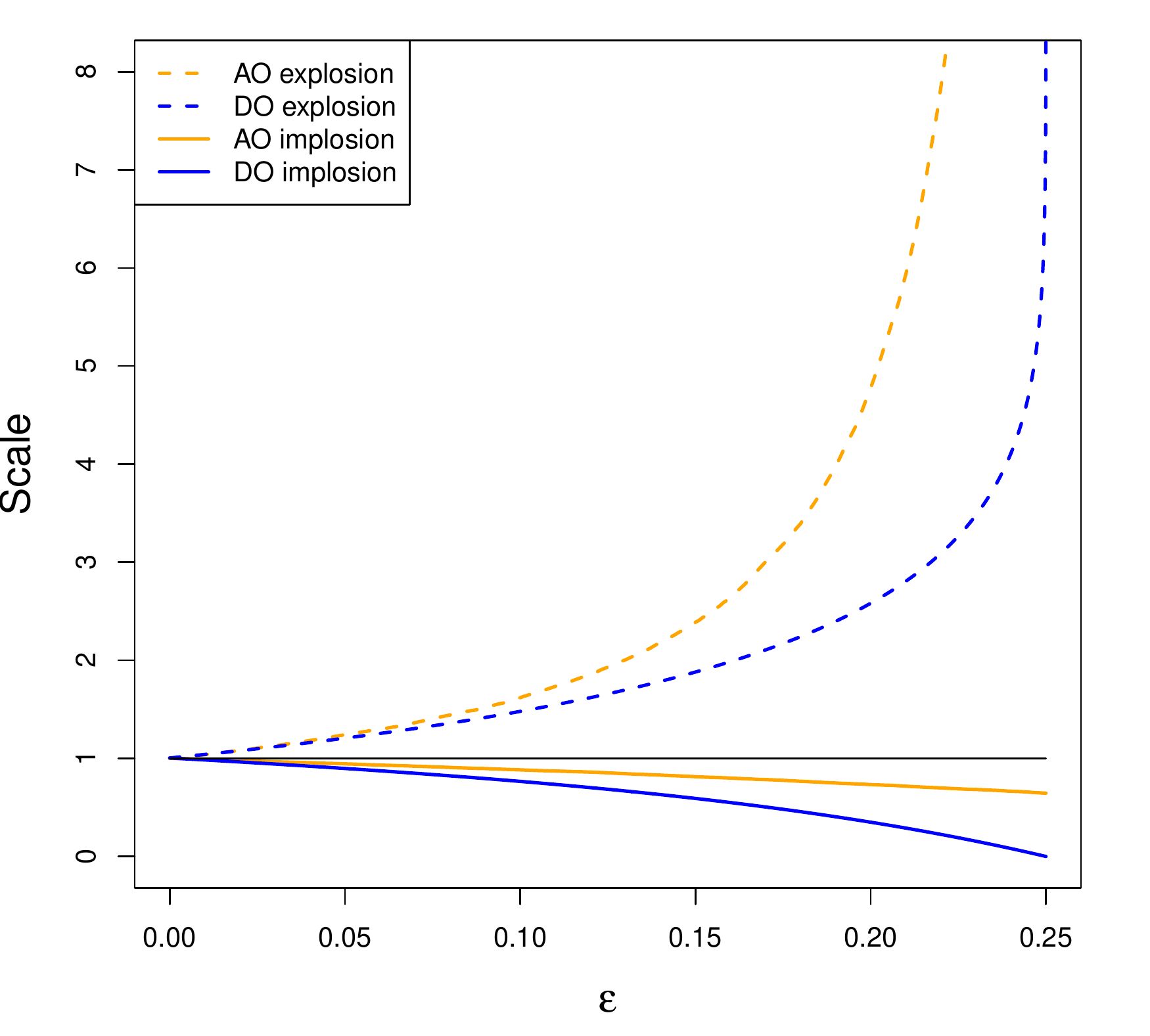}
\caption{Comparison of explosion and implosion bias of the 
         AO and DO scales.}
\label{fig:expimp}
\end{figure}

The blue curves in Figure~\ref{fig:expimp} are the explosion 
and implosion bias curves of $\sam$ when $F=\Phi$ is the 
standard gaussian distribution, and the tuning constant
in $\rho_c$ is $c=2.1$ corresponding to 85\% efficiency.
By affine equivariance the curves for $\sbm$ are exactly
the same, so these are the curves of both DO scales.
The orange curves correspond to explosion and implosion of
the scales used in the adjusted outlyingness AO
under the same contamination.
We see that the AO scale explodes faster, due to using the
medcouple in its definition.
The fact that the DO scale is typically smaller enables
the DO to attain larger values in outliers.

Another tool to measure the (non-)robustness of a
procedure is the influence function (IF).
Let $T$ be a statistical functional, and consider the
contaminated distribution
$F_{\varepsilon,z}=(1-\varepsilon)F+\varepsilon \Delta(z)$\;. 
The influence function of $T$ at $F$ is then given by
\begin{equation*}
\mbox{IF}(z,T,F)=\lim_{\varepsilon \to 0}
    \frac{T(F_{\varepsilon,z})-T(F)}{\varepsilon}
		= \frac{\partial}{\partial \varepsilon}
	  \;T(F_{\varepsilon,z})\bigg|_{\varepsilon=0}		
\end{equation*}
and basically describes how $T$ reacts to a small amount 
of contamination. 

This concept justifies our choice for the function $\rho_c$\;.
Indeed, the IF of a fully iterated M-estimator of scale with 
function $\rho$ is proportional to $\rho(z)-\beta$ with the
constant $\beta=\int_{-\infty}^{\infty}{\rho(x) dF(x)}$.
We use $\rho=\rho_c$ with $c=2.1$. 
It was shown in \cite{Hampel:IFapproach} that at $F=\Phi$
this $\rho_c$ yields the M-estimator with highest asymptotic
efficiency subject to an upper bound on the absolute value 
of its IF.

We will now derive the IF of the one-step M-estimator $\sam$
given by~\eqref{eq:funcsasb}.

\begin{proposition} The influence function of $\sam$ is 
  given by
\begin{equation*}
\begin{array}{ll}
  2 \alpha \;\frac{\sam(F)}{\soa^2(F)} 
   \;\mbox{IF}(z,\sam,F)=
  &\left\{\frac{2}{\soa(F)}\int_{\med(F)}^{\infty}{\rho_c 
  \left(\frac{x-\med(F)}{\soa(F)}\right)dF(x)}\right. \\
	&\left. -\int_{\med(F)}^{\infty}{x\rho_c '
	 \left(\frac{x-\med(F)}{\soa(F)}\right)dF(x)}\right. \\
  &\left. +\med(F)\int_{\med(F)}^{\infty}{\rho_c'
	 \left(\frac{x-\med(F)}{\soa(F)}\right)dF(x)} \right\}
	 \mbox{IF}(z,\soa,F)\\
  &-\left\{\int_{\med(F)}^{\infty}{\rho_c'\left(
	 \frac{x-\med(F)}{\soa(F)}\right)dF(x)}\right\}
	 \mbox{IF}(z,\med,F)\\
  &+\left\{\one_{[\med(F),\infty)}(z)\rho_c 
	 (z-\med(F)) - \alpha\right\}
\end{array}
\end{equation*}
where $\mbox{IF}(z,\soa,F)$
is the influence function of $\soa$.
\end{proposition}

\begin{figure}[!hb]
\centering
\includegraphics[width=0.5\textwidth]{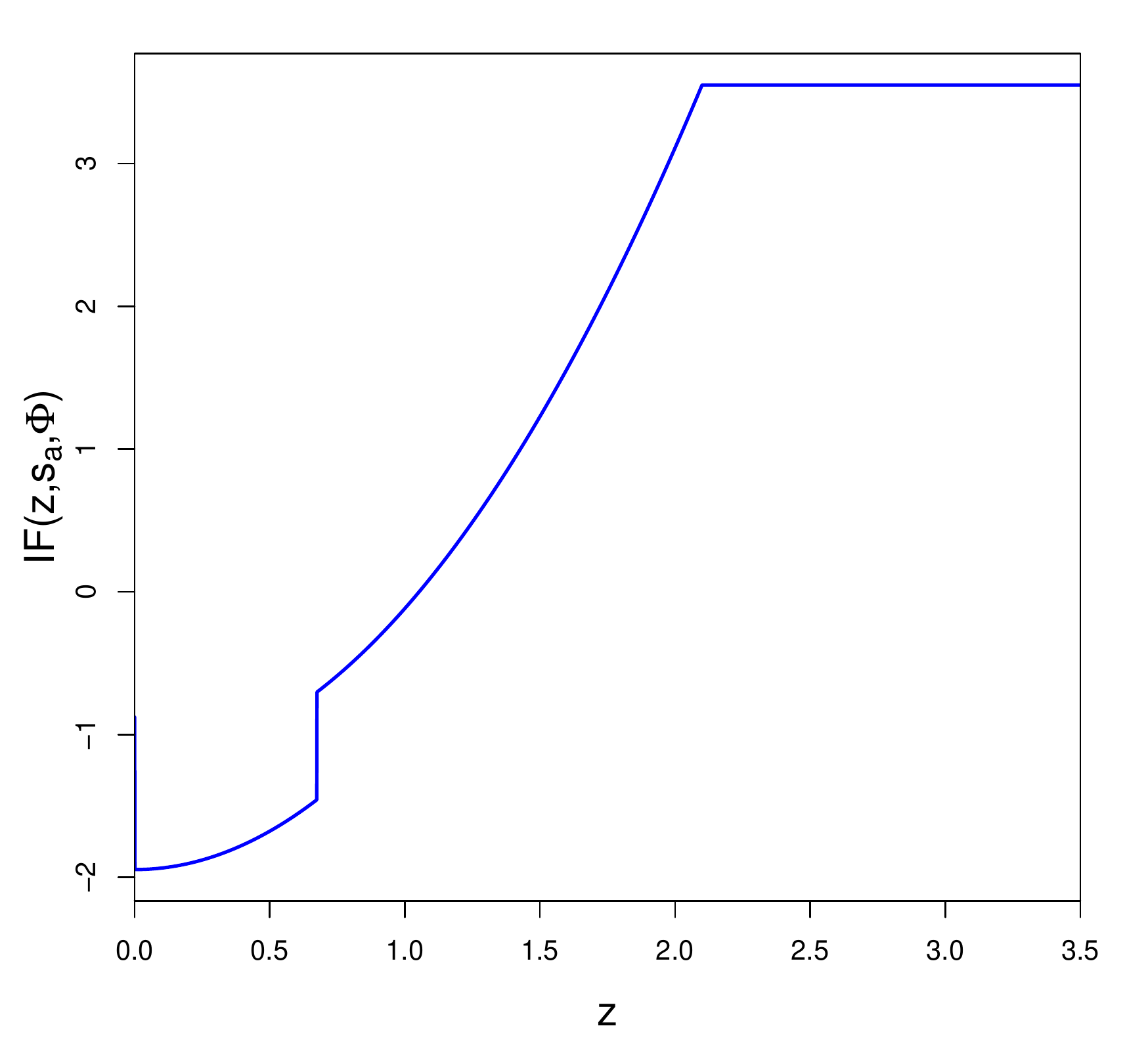}
\caption{Influence function of $\sam$ at $F=\Phi$\;.}
		\label{fig:IF_sa}
\end{figure}

The resulting IF of $\sam$ at $F=\Phi$ is shown in 
Figure \ref{fig:IF_sa}.
[Note that $\mbox{IF}(z,\sbm,\Phi)=\mbox{IF}(-z,\sam,\Phi)$.]
It is bounded, indicating that $s_a$ is robust to a small
amount of contamination even when it is far away.
Note that the IF has a jump at the third quartile 
$Q_3 \approx 0.674$ due to the initial estimate $\soa$.
If we were to iterate~\eqref{eq:funcsasb} to convergence
this jump would vanish, but then the explosion bias
would go 
up a lot, similarly to the computation
in~\cite{Rousseeuw:KstepM}.

Let us now compute the influence function of $\DO(x;F)$ 
given by~\eqref{eq:DO_univ}
for contamination in the point $z$, noting that $x$
and $z$ need not be the same.

\begin{proposition}
When $x>\med(F)$ it holds that
\begin{equation*}
  \mbox{IF}(z,DO(x),F) = \frac{-1}{\sam^2(F)}
  \left\{\mbox{IF}(z,\med,F)\sam(F)+
	\mbox{IF}(z,\sam,F)(x-\med(F))\right\}
\end{equation*}
whereas for $x<med(F)$ we obtain
\begin{equation*}
  \mbox{IF}(z,DO(x),F)= \frac{1}{\sbm^2(F)}
  \left\{\mbox{IF}(z,\med,F)\sbm(F) -
	\mbox{IF}(z,\sbm,F)(\med(F)-x)\right\}
  \;\;.
\end{equation*}
\end{proposition}
\begin{figure}[!ht]
\centering
\includegraphics[width=0.35\textwidth]{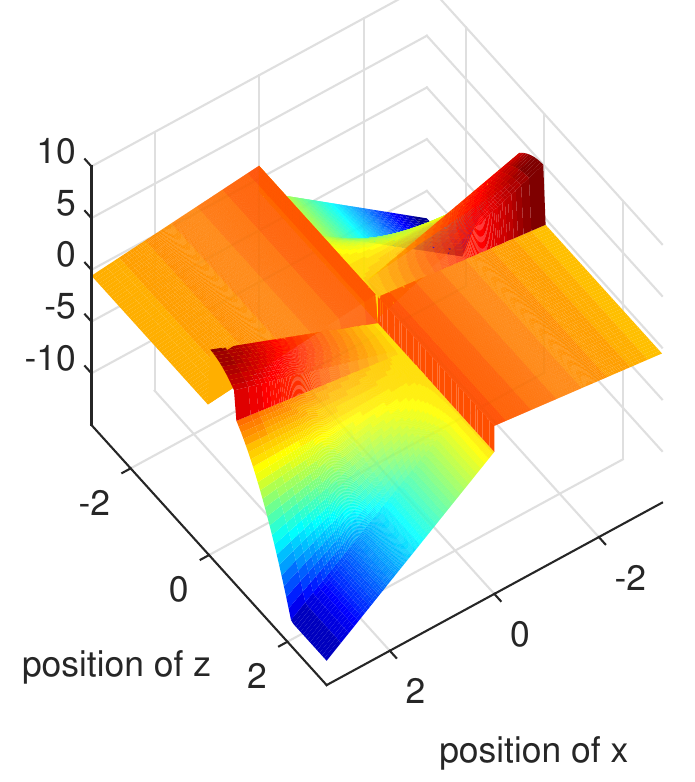}
\hspace{0.05\textwidth}
\includegraphics[width=0.56\textwidth]{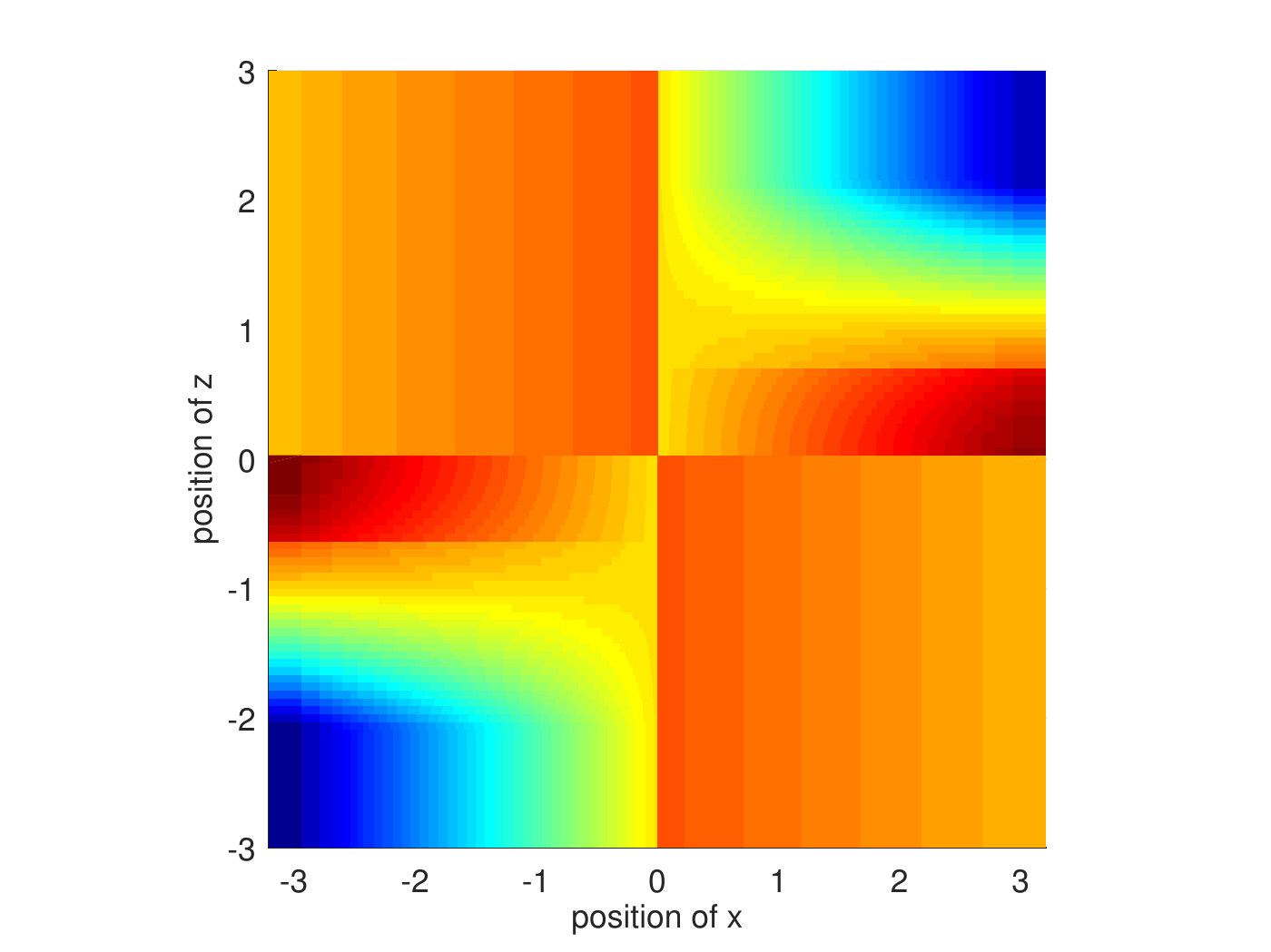}\\
\caption{Influence function of $\mbox{DO}(x)$ for $F=\Phi$.
  Left: 3D,
	right: 2D seen from above.}
		\label{fig:IF_DO}
\end{figure}

For a fixed value of $x$ the influence function of 
$\mbox{DO}(x)$ is bounded in $z$.
This is a desirable robustness property.
Figure \ref{fig:IF_DO} shows the influence function 
(which is a surface) when $F$ is the standard gaussian
distribution.

\subsection{Multivariate Setting}
In the multivariate setting we can apply the principle that a 
point is outlying with respect to a dataset if it stands out 
in at least one direction.
Like the Stahel-Donoho outlyingness, the multivariate DO 
is defined by means of univariate projections. 
To be precise, the DO of a d-variate point $\by$ relative 
to a $d$-variate sample $\bY=\{\by_1,\ldots,\by_n\}$ is 
defined as
\begin{equation}\label{eq:multivDO}
  \mbox{DO}(\by;\bY)= \sup\limits_{\bv \in \R^d}
	                      \mbox{DO}(\by^T\bv;\bY^T\bv)
\end{equation}
where the right hand side uses the univariate DO
of~\eqref{eq:DO_univ}.

To compute the multivariate DO
we have to rely 
on approximate algorithms, as it is impossible to project 
on {\it all} directions $\bv$ in $d$-dimensional space. 
A popular procedure to generate a direction is to randomly
draw $d$ data points, compute the hyperplane passing through
them, and then to take the direction $\bv$ orthogonal to it.
This guarantees that the multivariate DO is affine invariant. 
That is, the DO does not change when we add a constant vector 
to the data, or multiply the data by a nonsingular 
$d \times d$ matrix.

\begin{figure}[!hb]
\centering
\includegraphics[width=1.0\textwidth]
                {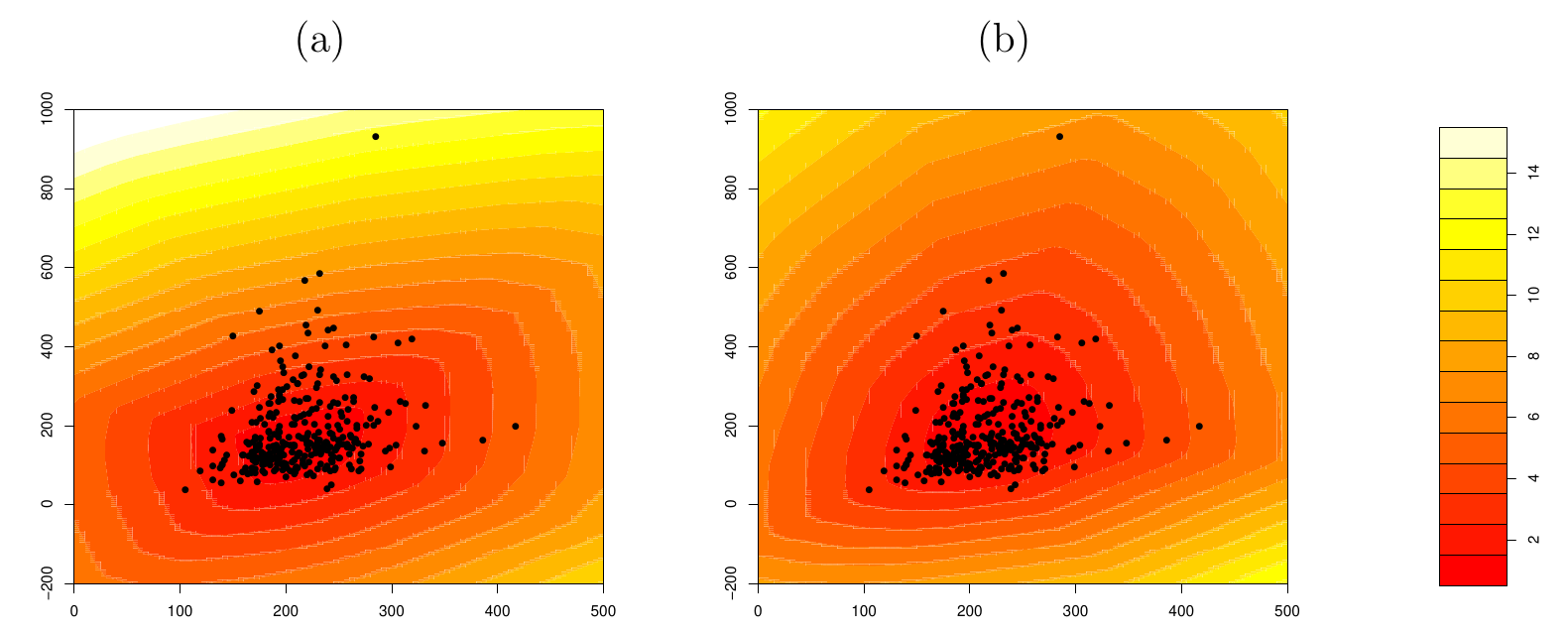}
\caption{Bloodfat data with (a) SDO contours, and (b)
         DO contours.}
\label{fig:bloodfat}
\end{figure}

As an illustration we take the bloodfat data set, which 
contains plasma cholesterol and plasma triglyceride 
concentrations (in mg/dl) of 320 male subjects for whom 
there is evidence of narrowing arteries
\citep{Hand:SmallData}. 
Here $n=320$ and $d=2$, and following 
\cite{Hubert:OutlierSkewed} we generated $250d = 500$ 
directions $\bv$.
Figure \ref{fig:bloodfat} shows the contour plots of both 
the DO and SDO measures. 
Their contours are always convex.
We see that the contours of the DO 
capture the skewness in the dataset, whereas those of the
SDO are more symmetric even though the data themselves
are not.

\subsection{Functional Directional Outlyingness}
\begin{figure}[!htb]
\centering
\includegraphics[width=0.5\textwidth]{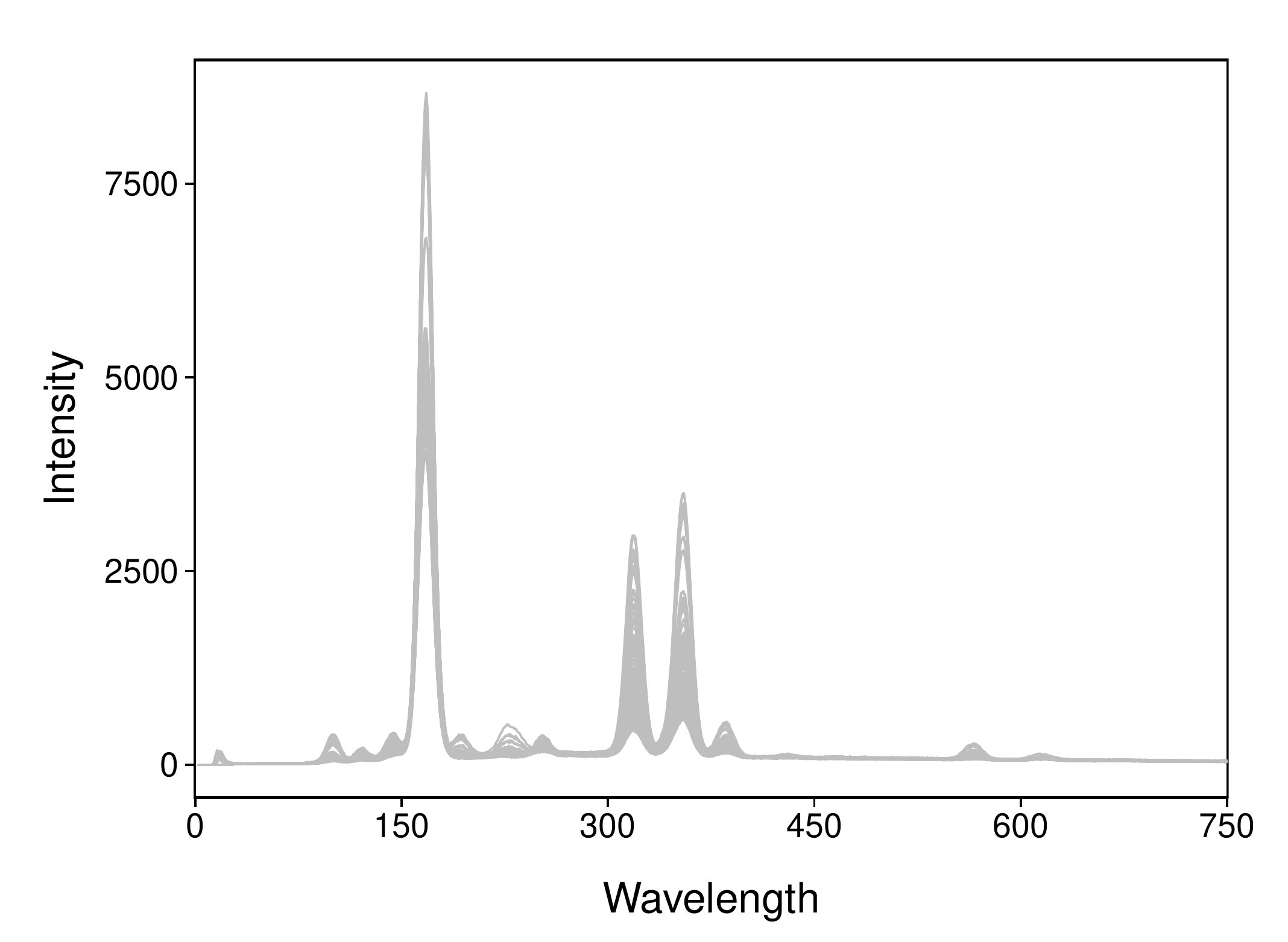}
\caption{Spectra of 180 archeological glass samples.}
		\label{fig:glassdata}
\end{figure}

We now extend the DO to data where the objects are functions.
To fix ideas we will consider an example.
The glass data set consists of spectra with $d = 750$ 
wavelengths resulting from spectroscopy on n = 180 
archeological glass samples \citep{Lemberge:PLS}. 
Figure \ref{fig:glassdata} shows the 180 curves.

At each wavelength the response is a single number, the 
intensity, so this is a univariate functional dataset.
However, we can incorporate the dynamic behavior of these
curves by numerically computing their first derivative.
This yields bivariate functions, where the response consists
of both the intensity and its derivative.

In general we write a functional dataset as
$\bY=\{Y_1,Y_2,\ldots,Y_n\}$ where each $Y_i$ is a 
$d$-dimensional function.
As in this example, the $Y_i$ are typically observed on a 
discrete set of points in their domain. 
For a univariate domain this set is denoted as
$\{t_1,\ldots,t_T\}$.

Now we want to define the DO of a d-variate function $X$
on the same domain, where $X$ need not be one of the $Y_i$\;.
For this we look at all the domain points $t_j$ in turn,
and define the
{\it functional directional outlyingness} (fDO) of $X$ with 
respect to the sample $\bm{Y}$ as
\begin{equation} \label{eq:fDO}
  \text{fDO}(X; \bY) = \sum_{j=1}^{T}{\DO(X(t_j);
	                     \bY(t_j)) \; W(t_j) }
\end{equation}
where $W(.)$ is a weight function for which 
$\sum_{j=1}^{T}{W(t_j)}=1$. 
Such a weight function allows us to assign a different 
importance to the outlyingness of a curve at different 
domain points. 
For instance, one could downweight time points near the 
boundaries if measurements are recorded less accurately at 
the beginning and the end of the process.   

\begin{figure}[!htb]
\centering
\includegraphics[width=0.5\textwidth]
                {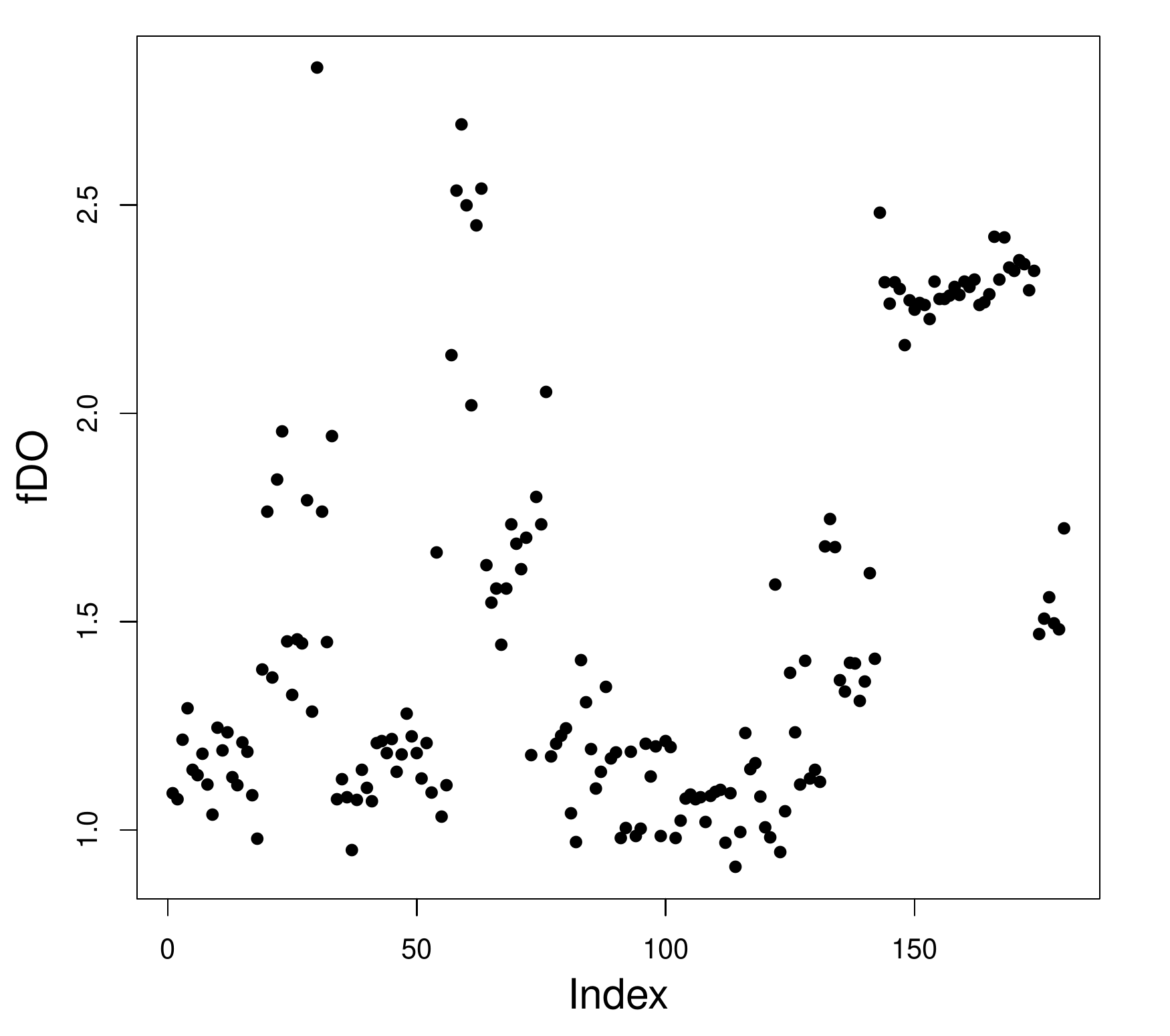}
\caption{fDO values of the 180 glass data functions.}
		\label{fig:fdo_glass}
\end{figure}

Figure~\ref{fig:fdo_glass} shows the fDO of the 180 
bivariate functions in the glass data, where $W(.)$ was set 
to zero for the first 13 wavelengths where the spectra
had no variability, and constant at the remaining wavelengths.
These fDO values allow us to rank the spectra from most to 
least outlying, but do not contain much information about 
how the outlying curves are different from the majority. 

\begin{figure}[!htb]
\centering
\includegraphics[width=\textwidth]
                {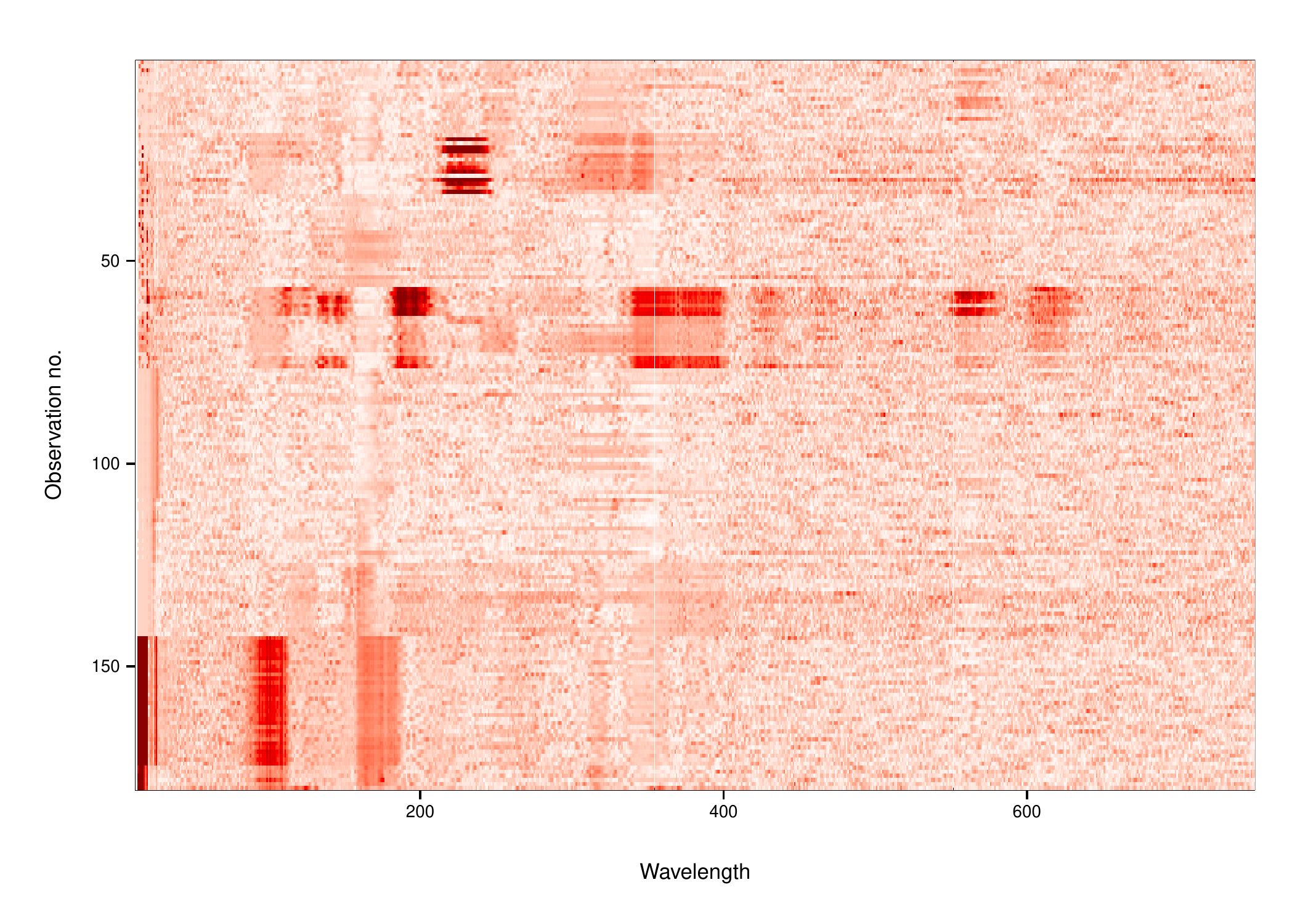}
\caption{Heatmap of DO of the glass data}
		\label{fig:DOmap_glass}
\end{figure}

In addition to this global outlyingness measure fDO we
also want to look at the local outlyingness.
To this end Figure~\ref{fig:DOmap_glass} shows the 
individual $\DO(Y_i(t_j);\bY(t_j))$ 
for each of the 180 functions $Y_i$ of the glass data
at each wavelength $t_j$.
Higher values of DO are shown by darker red in this heatmap.
Now we see that there are a few groups of curves with 
particular anomalies: one group around function 25, one 
around function 60, and one with functions near
the bottom.
Note that the global outlyingness measure fDO flags
outlying rows in this heatmap, whereas the dark spots
inside the heatmap can be seen as outlying cells.
It is also possible to sort the rows of the heatmap
according to their fDO values.
Note that the  
wavelength at which a dark spot in the
heatmap occurs allows to identify the chemical
element responsible.

As in \citep{Hubert:MFOD} we can transform the DO to the
multivariate depth function $1/(\mbox{DO}+1)$, and the fDO 
to the functional depth function $1/(\mbox{fDO}+1)$.

\section{Outlier Detection}
\label{sec:outlierdet}
\subsection{A Cutoff for Directional Outlyingness}
When analyzing a real data set we do not know its 
underlying distribution, but still we would like
a rough indication of which observations should
be flagged as outliers.
For this purpose we need an approximate cutoff value
on the DO.
We first consider non-functional data, leaving the
functional case for the next subsection.
Let $\bY = \{\by_1,\ldots,\by_n\}$ be a $d$-variate dataset 
($d \geqslant 1$) with directional outlyingness 
values $\{\DO_1,\ldots,\DO_n\}$. 
The $\DO_i$ have a right-skewed distribution, so we
transform them to
   $\{\LDO_1,\ldots,\LDO_n\}=
    \{\log(0.1+\DO_1),\ldots,\log(0.1+\DO_n)\}$
of which the majority is closer to gaussian.
Then we center and normalize the resulting values in a 
robust way and compare them to a high gaussian quantile.
For instance, we can flag $\by_i$ as outlying whenever 
\begin{equation}\label{eq:cutoffDO}
  \frac{\text{LDO}_i-\med(\text{LDO})}
	{\text{MAD}(\text{LDO})} > \Phi^{-1}(0.995) \, . 
\end{equation}
so the cutoff for the 
DO values is
$c =\mbox{exp} \big( \med(LDO)+\MAD(LDO)\; 
            \Phi^{-1}(0.995) \big) - 0.1\;$.
(Note that we can use the same formulas for functional
data by replacing DO by fDO.)

\begin{figure}[!htb]
\centering
\includegraphics[width=0.6\textwidth]
                {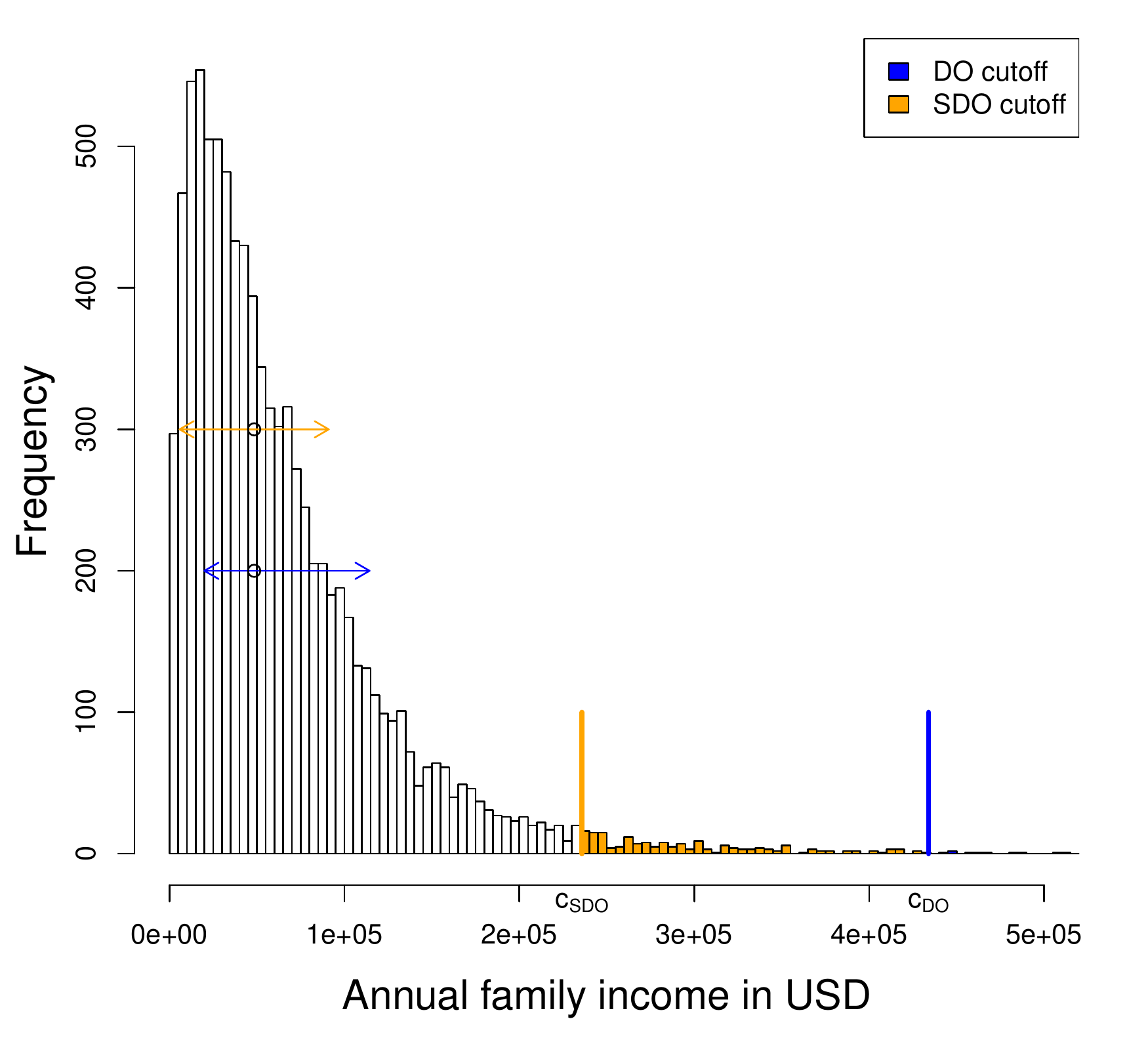}
\caption{Outlier cutoffs for the family income data.}
		\label{fig:scalesfill}
\end{figure}

For an illustration we return to the family income data 
of Figure~\ref{fig:scales}.
The blue vertical line in Figure~\ref{fig:scalesfill}
corresponds to the DO cutoff, whereas the orange line is
the result of the same computation applied to the
Stahel-Donoho outlyingness.
The DO cutoff is the more conservative one, because it takes 
the skewness of the income distribution into account.

\begin{figure}[!htb]
\centering
\includegraphics[width=0.5\textwidth]
                {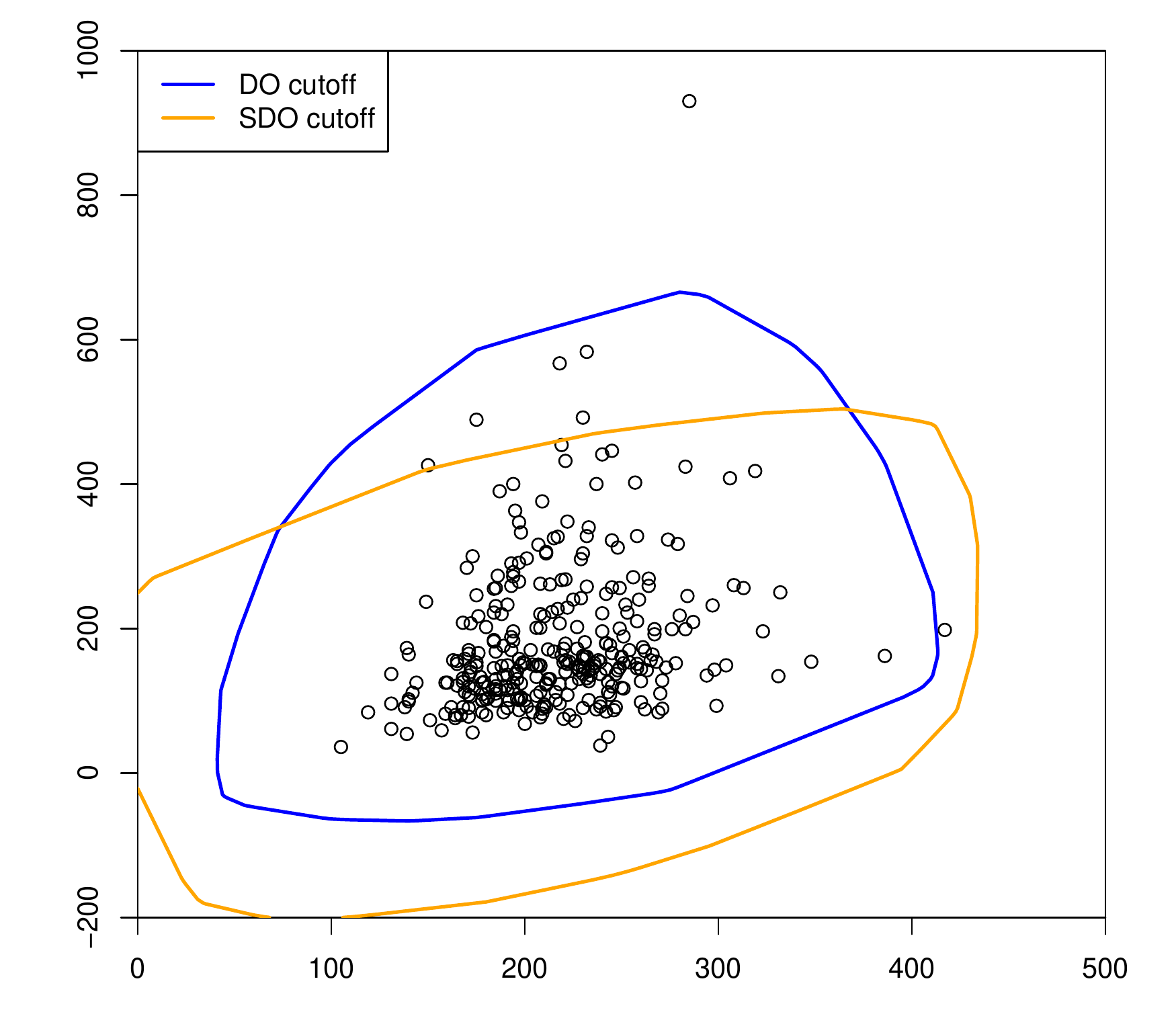}
\caption{Outlier detection on bloodfat data}
		\label{fig:bloodfat_outlier}
\end{figure}

Figure~\ref{fig:bloodfat_outlier} shows the DO and SDO 
cutoffs for the bivariate bloodfat data of 
Figure~\ref{fig:bloodfat}.
The DO captures the skewness in the data and flags only
two points as outlying, whereas the SDO takes a more 
symmetric view and also flags five of the presumed inliers.

\subsection{The Functional Outlier Map}
When the data set consists of functions there can be
several types of outlyingness.
As an aid to distinguish between them, 
\cite{Hubert:MFOD} introduced a graphical tool 
called the {\it functional outlier map} (FOM). 
Here we will extend the FOM to the new DO measure and add
a cutoff to it, in order to increase its utility.

Consider a functional dataset $\bY=\{Y_1,Y_2,\ldots,Y_n\}$.
The fDO [see~\eqref{eq:fDO}] of a function $Y_i$ can be 
interpreted as the `average outlyingness' of its (possibly
 multivariate) function values.
We now also measure the {\it variability} of its DO values, 
by
\begin{equation}\label{eq:vDO}
  \text{vDO}(Y_i;\bm{Y}) = 
	\frac{\text{stdev}_j\big( 
	\text{DO}(Y_i(t_j);\bY(t_j))\big)}
	{1+\text{fDO}(Y_i;\bY)} \;\;.
\end{equation}
Note that~\eqref{eq:vDO} has the fDO in the denominator in 
order to measure relative instead of absolute variability. 
This can be understood as follows. 
Suppose that the functions $Y_i$ are centered around zero 
and that $Y_k(t_j)=2 \; Y_i(t_j)$ for all $j$. 
Then $\text{stdev}_j(\text{DO}(Y_k(t_j);\bY(t_j))) =
   2\;\text{stdev}_j(\text{DO}(Y_i(t_j);\bY(t_j)))$ 
but their relative variability is the same. 
Because $\;\text{fDO}(Y_k;\bY)=2\;\text{fDO}(Y_i;\bY)$, 
putting fDO in the denominator normalizes for this.
In the numerator we could also compute a weighted standard
deviation with the weights $W(t_j)$ from~\eqref{eq:fDO}.

The FOM is then the scatter plot of the points
\begin{equation}\label{eq:fom}
  \left(\;\text{fDO}(Y_i;\bm{Y})\;,
    \;\text{vDO}(Y_i;\bm{Y})\;\right) 
\end{equation}
for $i=1,\ldots,n$.
Its goal is to reveal outliers in the data, and its 
interpretation is fairly straightforward. 
Points in the lower left part of the FOM represent regular 
functions which hold a central position in the data set. 
Points in the lower right part are functions with a high fDO 
but a low variability of DO values. 
This happens for {\it shift outliers}, i.e.\ functions 
which have a regular shape 
but are shifted on the whole domain. 
Points in the upper left part have a low fDO but a high vDO. 
Typical examples are local outliers, i.e.\ functions which 
only display outlyingness over a small part of their domain.
The points in the upper right part of the FOM have both a high 
fDO and a high vDO. These correspond to functions which are 
strongly outlying on a substantial part of their domain.

\begin{figure}[!htb]
\centering
\includegraphics[width=0.7\textwidth]
                {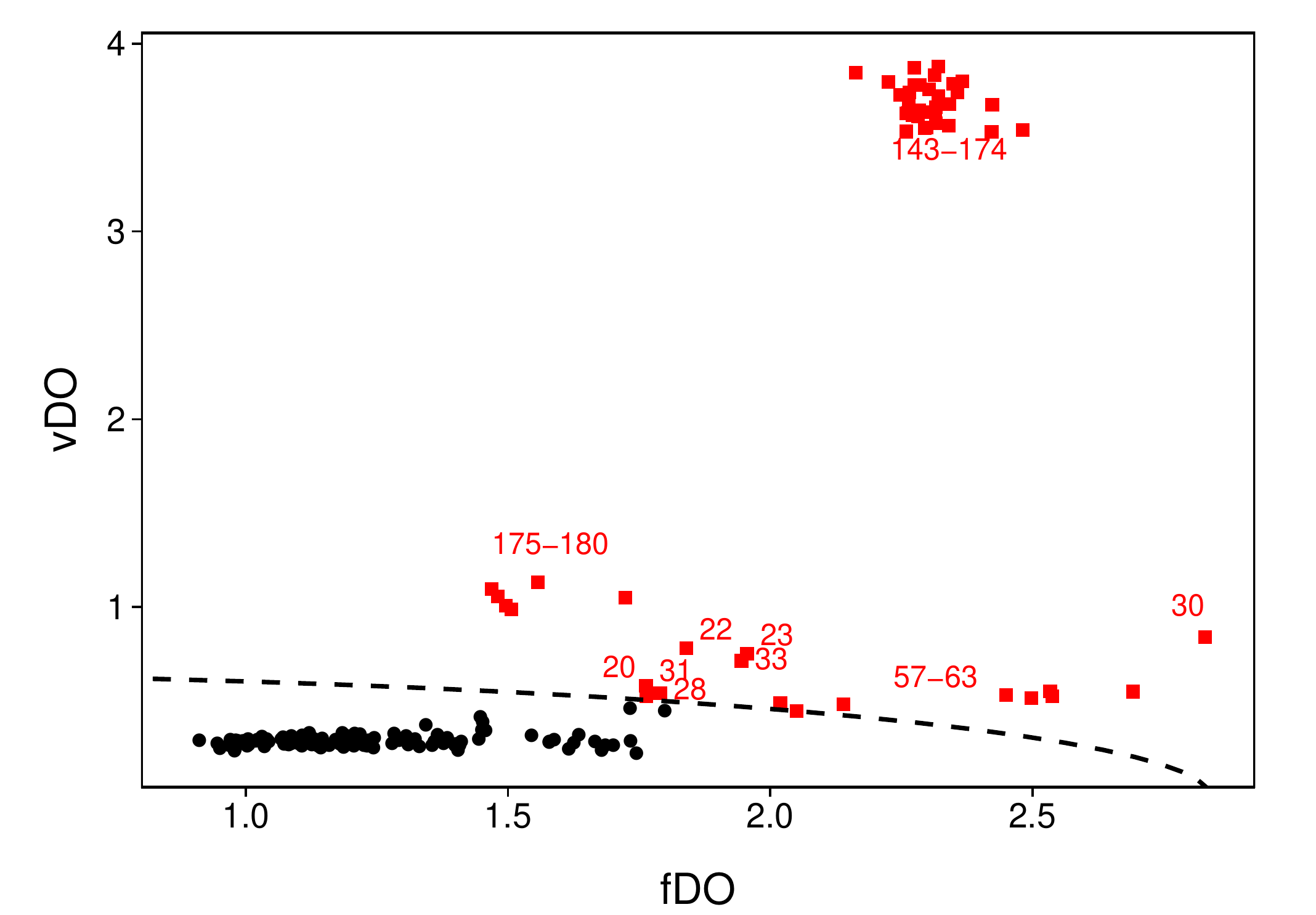}
\caption{FOM of the glass data}
		\label{fig:glass_fom}
\end{figure}

As an illustration we revisit the glass data.
Their FOM in Figure~\ref{fig:glass_fom} contains a lot more
information than their fDO values alone in 
Figure~\ref{fig:fdo_glass}.
In the heatmap (Figure~\ref{fig:DOmap_glass}) we noticed 
three groups of outliers, which also stand out in the FOM.
The first group consists of the spectra 20, 22, 23, 28, 30, 
31 and 33.
Among these, number 30 lies furthest to the right in the FOM.
It corresponds to row 30 in Figure~\ref{fig:DOmap_glass} 
which has a dark red piece.
It does not look like a shift outlier, for which the row
would have a more homogeneous color (hence a lower vDO).
The second group, with functions 57--63, occupies a similar
position in the FOM. 
The group standing out the most consists of functions 
143--174. 
They are situated in the upper part of the FOM, indicating 
that they are shape outliers.
Indeed, they deviate strongly from the majority in three 
fairly small series of wavelengths. 
Their outlyingness is thus more local than that 
of functions 57--63.

We now add a new feature to the FOM, namely a rule to 
flag outliers. 
For this we define 
the \textit{combined functional outlyingness} (CFO) of a 
function $Y_i$ as
\begin{equation} \label{eq:cfo}
  \text{CFO}_i=\text{CFO}(Y_i; \bm{Y}) = \sqrt{(\text{fDO}_i
	/\med(\text{fDO}))^2+(\text{vDO}_i/\med(\text{vDO}))^2}
\end{equation}
where\; $\text{fDO}_i = \text{fDO}(Y_i;\bm{Y})$ and\;  
$\med(\text{fDO}) = \med(\text{fDO}_1,\ldots,\text{fDO}_n)$, 
and similarly for vDO. 
Note that the CFO characterizes the points in the FOM through 
their Euclidean distance to the origin, after scaling. 
We expect outliers to have a large CFO. 
In general, the distribution of the CFO is unknown but 
skewed to the right. 
To construct a cutoff for CFO we use the same reasoning
as for the cutoff~\eqref{eq:cutoffDO} on fDO:
First we compute $\text{LCFO}_i = \log(0.1 + \text{CFO}_i)$ 
for all $i=1,\ldots,n$, and then we flag function $Y_i$ as 
outlying if
\begin{equation} \label{eq:cutoff_fom}
  \frac{\text{LCFO}_i-\med(\text{LCFO})}
	{\text{MAD}(\text{LCFO})} > \Phi^{-1}(0.995) \, .
\end{equation}
This yields the dashed curve (which is part of an ellipse) 
in the FOM of Figure~\ref{fig:glass_fom}.

\section{Application to Image Data}
Images are functions on a bivariate domain.
In practice the domain is a grid of discrete points, e.g.\ 
the horizontal and vertical pixels of an image. 
It is convenient to use two indices $j=1,\ldots,J$ and 
$k=1,\ldots,K$\;, one for each dimension of the grid, to 
characterize these points. 
An image (or a surface) is then a function on the
$J \times K$ points of the grid.
Note that the function values can be univariate, like
gray intensities, but they can also be multivariate, e.g.
the intensities of red, green and blue (RGB). 
In general we will write an image dataset as a sample
$\bY=\{Y_1,Y_2,\ldots,Y_n\}$ where each $Y_i$ is a function
from $\{(j,k);\;j=1,\ldots,J\;\mbox{and}\;k=1,\ldots,K\}$\;
to $\R^d$\;.

The fDO~\eqref{eq:fDO} and vDO~\eqref{eq:vDO} notions 
that we saw for functional data with a univariate domain 
can easily be extended to functions with a bivariate domain
by computing
\begin{equation}\label{eq:fDO_surface}
  \text{fDO}(Y_i; \bY) = \sum_{j=1}^{J}{\sum_{k=1}^{K}
	    {\DO(Y_i(j,k);\bY(j,k)) \; W_{jk}}}
\end{equation}
where the weights $W_{jk}$ must satisfy 
$\sum_{j=1}^J \sum_{k=1}^K W_{jk} = 1$, and
\begin{equation}\label{eq:vDO_surface}
  \text{vDO}(Y_i;\bY)=\frac{\text{stdev}_{j,k}
	      (\text{DO}(Y_i(j,k);\bY(j,k)))}
				{1+\text{fDO}(Y_i;\bY)}
\end{equation}
where the standard deviation can also be weighted by
the $W_{jk}$. 
(The simplest weight function is the constant $W_{jk}=1/(JK)$ 
for all $j=1,\ldots,J$ and $k=1,\ldots,K$.)
Note that~\eqref{eq:fDO_surface} and~\eqref{eq:vDO_surface} 
can trivially be extended to functions with domains in
more than 2 dimensions, such as three-dimensional images 
consisting of voxels.
In each case we obtain $\mbox{fDO}_i$ and $\mbox{vDO}_i$ 
values that we can plot in a FOM, with cutoff 
value~\eqref{eq:cutoff_fom}.

As an illustration we analyze a dataset containing MRI 
brain images of 416 subjects aged between 18 and 
96~\citep{Marcus:OASIS}, which can be freely accessed 
at {\it www.oasis-brains.org}\;.
For each subject several images are provided; we will use 
the masked atlas-registered gain field-corrected images 
resampled to 1mm isotropic pixels. 
The masking has set all non-brain pixels to an intensity 
value of zero. 
The provided images are already normalized, meaning that 
the size of the head is exactly the same in each image.  
The images have 176 by 208 pixels, with grayscale values 
between 0 and 255. 
All together we thus have 416 observed images $Y_i$ 
containing univariate intensity values $Y_i(j,k)$, where 
$j=1,\ldots,J=176$ and $k=1,\ldots,K=208$.

There is more information in such an image than just the 
raw values. 
We can incorporate shape information by computing 
the gradient in every pixel of the image. 
The gradient in pixel $(j,k)$ is defined as 
the 2-dimensional vector 
$\nabla Y_i(j,k) = \left(\frac{\partial Y_i(j,k)}
 {\partial j},
 \frac{\partial Y_i(j,k)}{\partial k} \right)$
in which the derivatives have to be approximated
numerically. 
In the pixels at the boundary of the brain we compute
forward and backward finite differences, and for
the other pixels we employ central differences.
In the horizontal direction we thus compute one of 
three expressions:
$$\frac{\partial Y_i(j,k)}{\partial j}=
\left\{
\begin{array}{ll}
  (-3 \, Y_i(j,k) + 4 \, Y_i(j+1,k) - Y_i(j+2,k))/2
	& \;\;(\text{forward difference})\\
  (Y_i(j+1,k)-Y_i(j-1,k))/2
  & \;\;(\text{central difference})\\
  (Y_i(j-2,k)-4 \, Y_i(j-1,k)+3\,  Y_i(j,k))/2
  & \;\;(\text{backward difference})\\
\end{array}
\right.$$
depending on where the pixel is located.
The derivatives in the vertical direction are computed 
analogously.

\begin{figure}[!htb]
\centering
\includegraphics[width=1.0\textwidth]
                {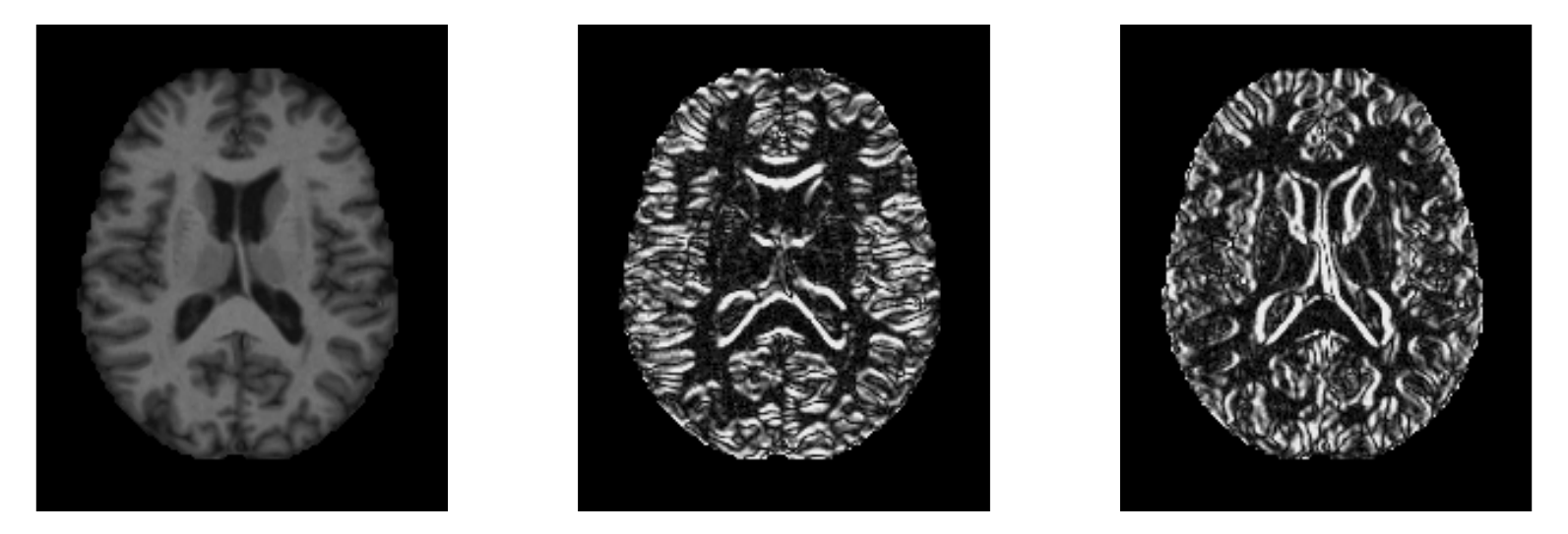}
\caption{Original MRI image of subject 387, and its 
         derivatives in the horizontal and vertical
				 direction.}
\label{fig:mridata}
\end{figure}

Incorporating these derivatives yields a dataset of 
dimensions $416 \times 176 \times 208 \times 3$, so the
final $Y_i(j,k)$ are trivariate. 
For each subject we thus have three data matrices which 
represent the original MRI image and its derivatives in 
both directions. 
Figure~\ref{fig:mridata} shows these three matrices for 
subject number 387.

The functional DO of an MRI image $Y_i$ 
is given by~\eqref{eq:fDO_surface}:
\begin{equation*}
\text{fDO}(Y_i;\bY) = \frac{1}{176 \times 208}
           \sum_{j=1}^{176}\sum_{k=1}^{208} 
		       \DO(Y_i(j,k); \bY(j,k))\; W_{jk}\; ,
\end{equation*}
where $\DO(Y_i(j,k);\bY(j,k))$ is the DO of the
trivariate point $Y_i(j,k)$ relative to the trivariate
dataset \;$\{Y_1(j,k),\ldots,Y_{416}(j,k)\}$\;.
In this example the have set the weight $W_{jk}$ equal to 
zero at the grid points that are not part of the brain, 
shown as the black pixels around it.
The grid points inside the brain receive full weight.

\begin{figure}[!htb]
\centering 
\includegraphics[width=0.75\textwidth]
                {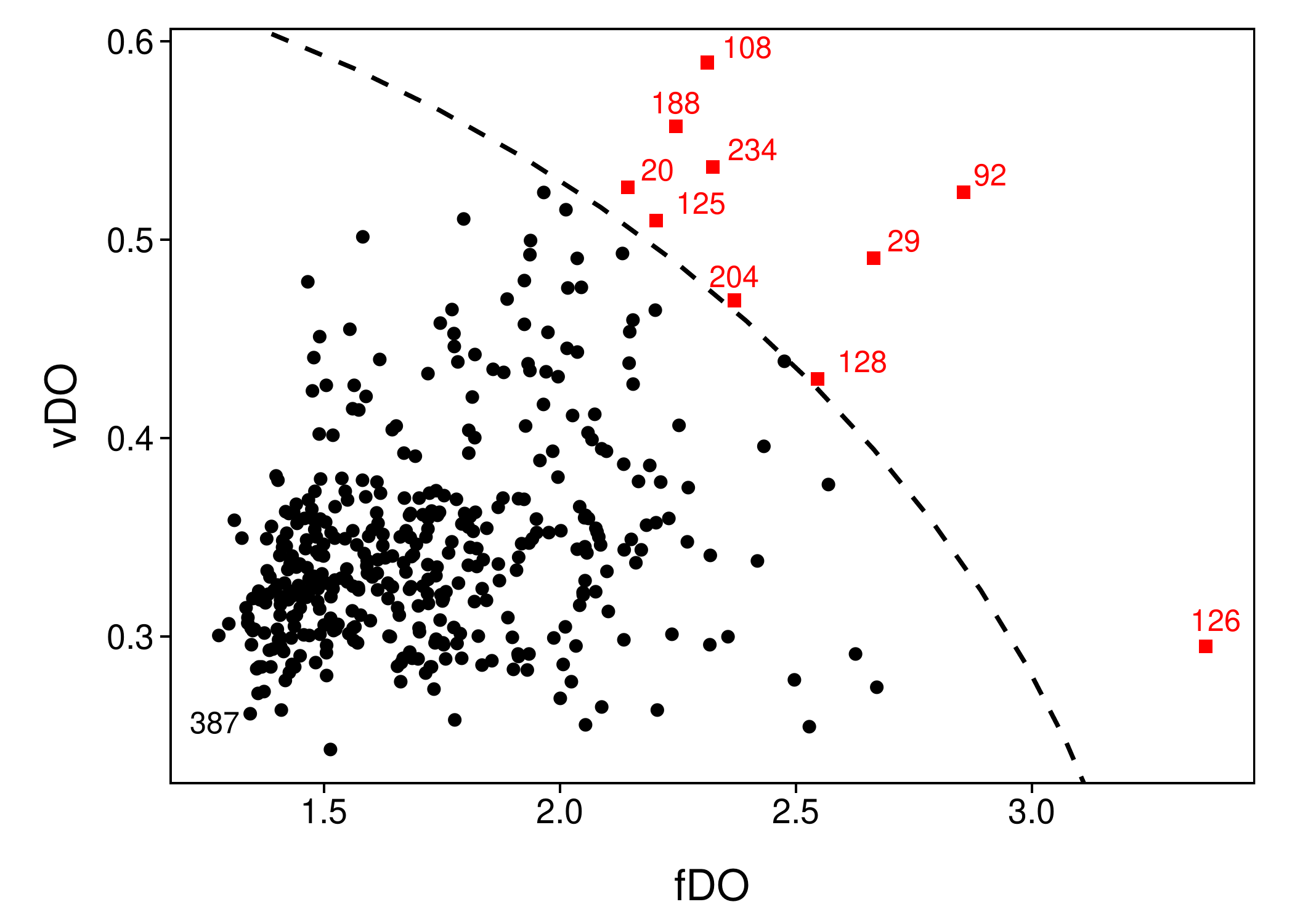} 
\caption{FOM of the MRI dataset.}
\label{fig:mrifom}
\end{figure}

Figure~\ref{fig:mrifom} shows the resulting FOM, which
indicates the presence of several outliers.
Image 126 has the highest fDO combined with a 
relatively low vDO. 
This suggests a shift outlier, i.e.\ a function whose values 
are all shifted relative to the majority of the data. 
Images 29 and 92 have a large fDO in combination with a high 
vDO, indicating that they have strongly outlying subdomains. 
Images 108, 188 and 234 have an fDO which is on the high end
relative to the dataset, which by itself does not make them 
outlying. 
Only in combination with their large vDO are they flagged as 
outliers. 
These images have strongly outlying subdomains which are 
smaller that those of functions 29 and 92.
The remaining flagged images are fairly close to the 
cutoff, meaning they are merely borderline cases.

\begin{figure}[!ht]
\centering
\includegraphics[width=0.75\textwidth]
                {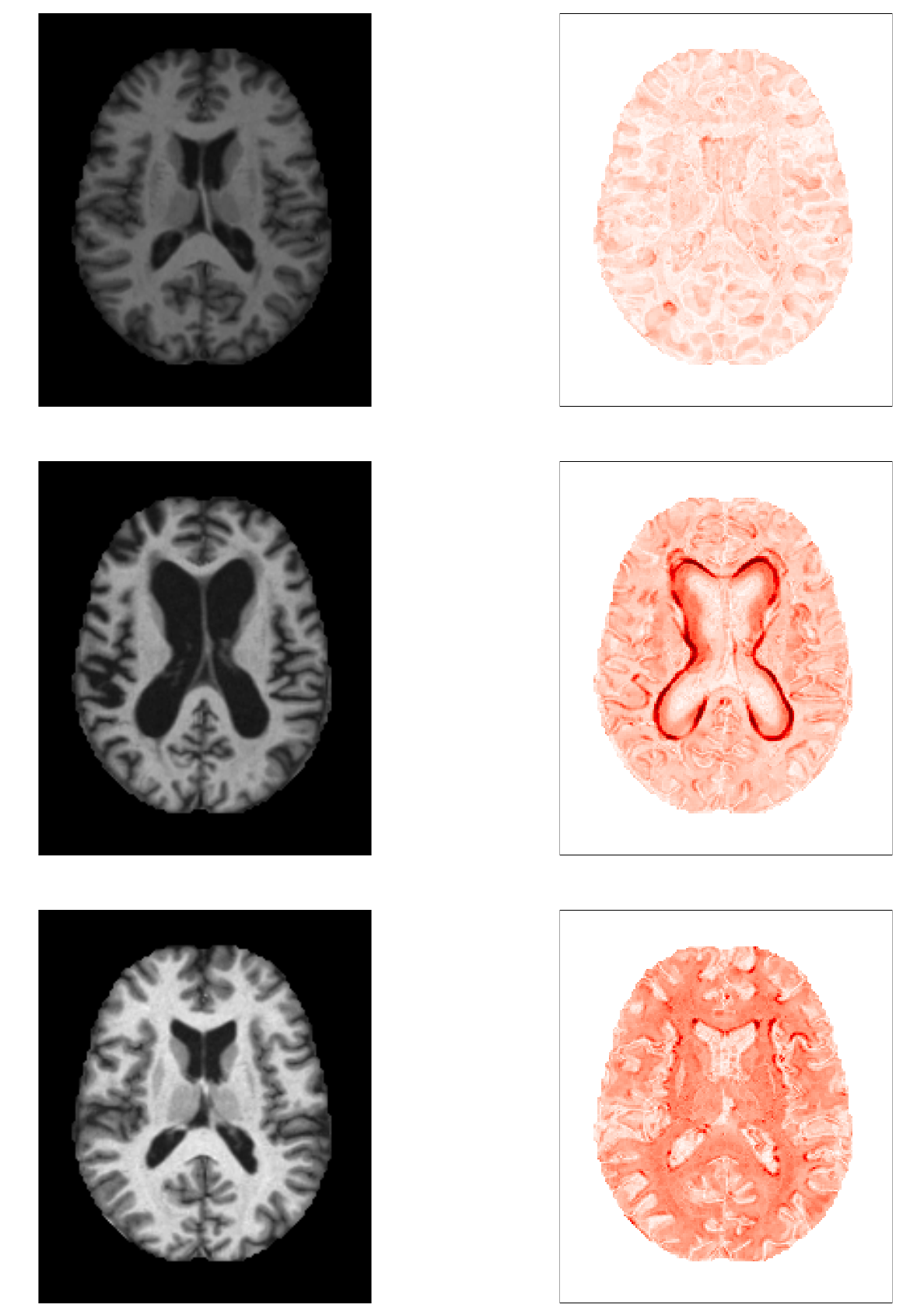} 
\caption{MRI image (left) and DO heatmap (right) of subjects
         387 (top), 92 (middle), and 126 (bottom).}
\label{fig:mrigrid}
\end{figure}

In order to find out why a particular image is outlying it 
is instructive to look at a heatmap of its DO values.
In Figure~\ref{fig:mrigrid} we compare the MRI images 
(on the left) and the DO heatmaps (on the right) of subjects 
387, 92, and 126. 
DO values of 15 or higher received the darkest color.
Image 387 has the smallest CFO value, and can be 
thought of as the most central image in the dataset. 
As expected, the DO heatmap of image 387 shows very few 
outlying pixels. 
For subject 92, the DO heatmap nicely marks the region in 
which the MRI image deviates most from the majority of 
the images. 
Note that the boundaries of this region have the highest outlyingness. 
This is due to including the derivatives in the analysis, 
as they emphasize the pixels at which the grayscale intensity
changes. 
The DO heatmap of subject 126 does not show any extremely 
outlying region but has a rather high outlyingness over 
the whole domain, which explains its large fDO and regular 
vDO value. 
The actual MRI image to its left is globally lighter than 
the others, confirming that it is a shift outlier.

\section{Application to Video}
We analyze a surveillance video of a beach, filmed with a 
static camera~\citep{Li:ObjectDetection}. 
This dataset can be found at
{\it http://perception.i2r.a-star.edu.sg/bk\_model/bk\_index.html} 
and consists of 633 frames.

\begin{figure}[!htb]
\centering
\vspace{1ex}
\includegraphics[width=1.0\textwidth]
                {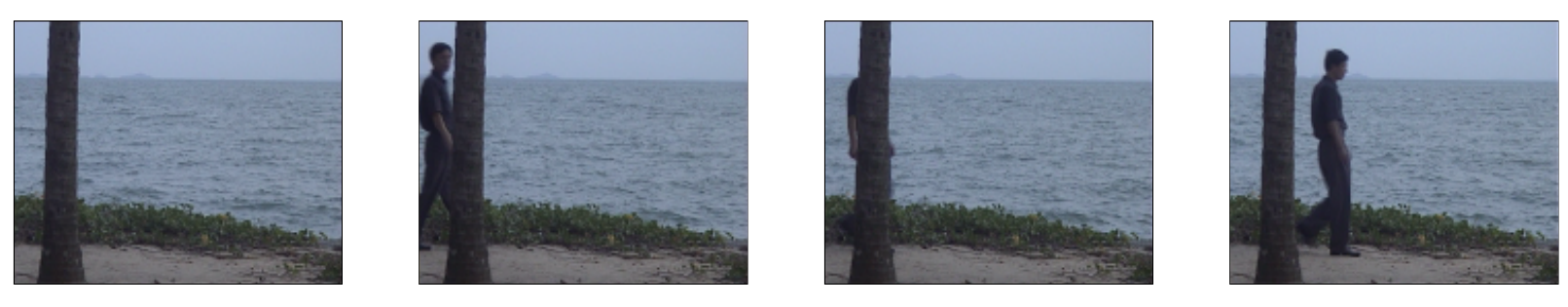}
\caption{Frames number 100, 487, 491 and 500 from the video dataset}
\label{fig:videodata}
\end{figure}

The first 8 seconds of the video show a beach with a tree,
as in the leftmost panel of Figure~\ref{fig:videodata}. 
Then a man enters the screen from the left (second panel), 
disappears behind the tree (third panel), and then 
reappears to the right of the tree and stays on 
screen until the end of the video. 
The frames have $160 \times 128$ pixels and are stored 
using the RGB (Red, Green and Blue) color model, so each 
frame corresponds to three matrices of size
$160 \times 128$. 
Overall we have 633 frames $Y_i$ containing trivariate 
$Y_i(j,k)$ for $j=1,\ldots,J=160$ and $k=1,\ldots,K=128$. 

\begin{figure}[!htb]
\centering
\includegraphics[width=0.8\textwidth]
                {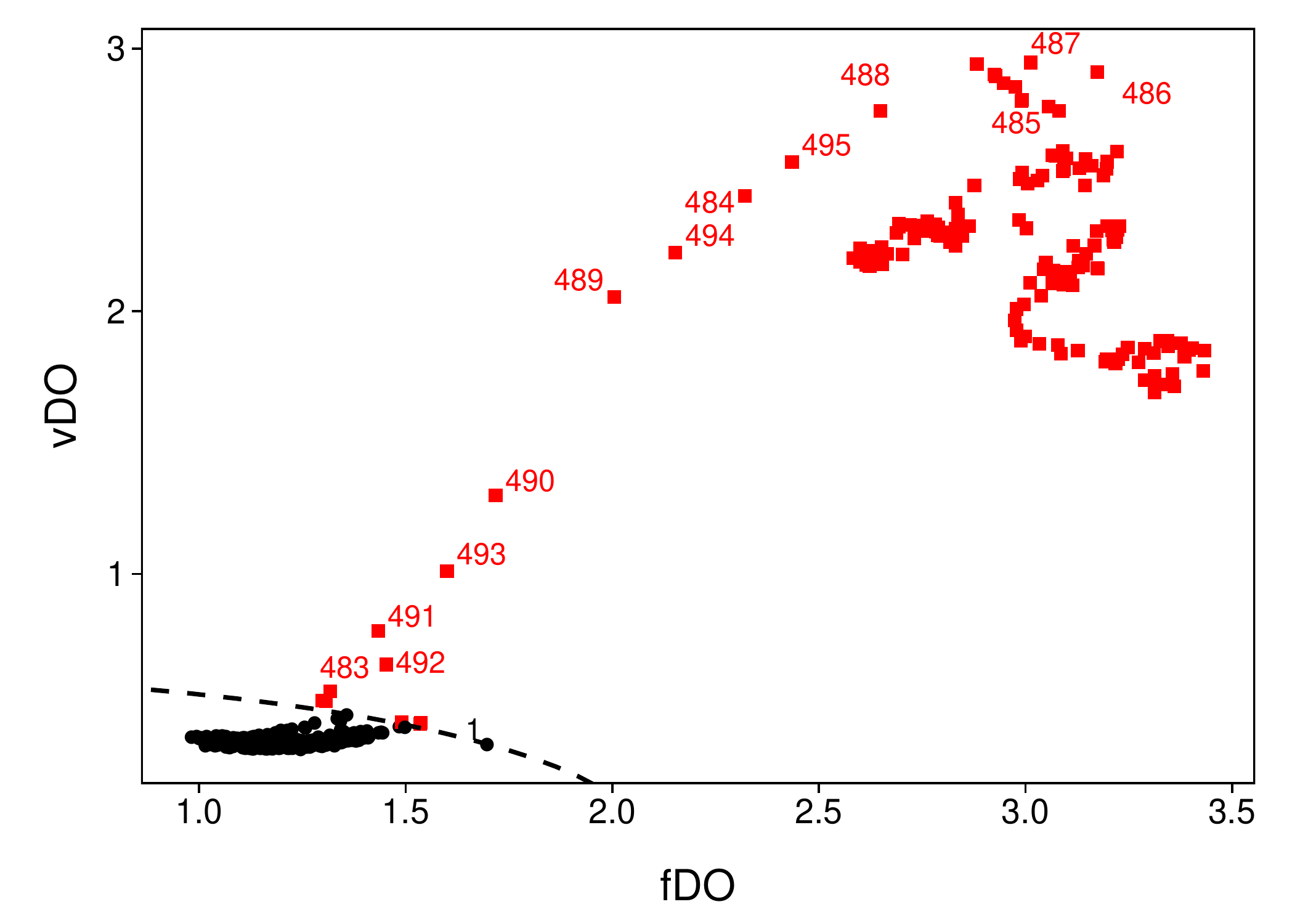}
\caption{FOM of the video data.}
		\label{fig:video_fom}
\end{figure}

Computing the fDO~\eqref{eq:fDO_surface} in this data set
is time consuming since we have to execute the projection 
pursuit algorithm~\eqref{eq:multivDO} in $\R^3$ for each
pixel, so $160 \times 128 = 20,480$ times. 
The entire computation took about on hour and a half on 
a laptop.
Therefore we switch to an alternative computation.
We define the componentwise DO of a $d$-variate 
point $\by$ relative to a $d$-variate 
sample $\bY=\{\by_1,\ldots,\by_n\}$ as
\begin{equation}\label{eq:CDO}
  \mbox{CDO}(\by;\bY)= 
	   \sqrt{\sum_{h=1}^{d} \DO(y_h;Y_h)^2}
\end{equation}
where $\DO(y_h;Y_h)$ is the univariate DO of the
$h$-th coordinate of $\by$ relative to the $h$-th 
coordinate of $\bY$.
Analyzing the video data with this componentwise procedure 
took under 2 minutes, so it is about 50 times faster than
with projection pursuit, and it produced almost the same 
FOM. 
Figure~\ref{fig:video_fom} shows the FOM obtained from the 
CDO computation.

The first 480 frames, which depict the beach and the tree 
with only the ocean surface moving slightly, are found at 
the bottom left part of the FOM. 
They fall inside the dashed curve that separates the regular
frames from the outliers.
At frame 483 the man enters the picture, making the 
standard deviation of the DO rise slightly. 
The fDO increases more slowly, as the fraction of the pixels 
covered by the man is still low at this stage.
This frame can thus be seen as locally outlying. 
The subsequent frames 484--487 have very high fDO and vDO. 
In them the man is clearly visible between the left border 
of the frame and the tree, so these frames have outlying 
pixels in a substantial part of their domain. 
Frames 489--492 see the man disappear behind the tree, 
so the fDO goes down as the fraction of outlying
pixels decreases.
From frame 493 onward the man reappears to the right of the 
tree and stays on screen until the end.
These frames contain many outlying pixels, yielding points 
in the upper right part of the FOM.

In the FOM we also labeled frame 1, which lies close to 
the outlyingness border. 
Further inspection indicated that this frame is a bit 
lighter than the others, which might be due to the 
initialization of the camera at the start of the video.

\begin{figure}[!ht]
\centering
\includegraphics[width=0.99\textwidth]
                {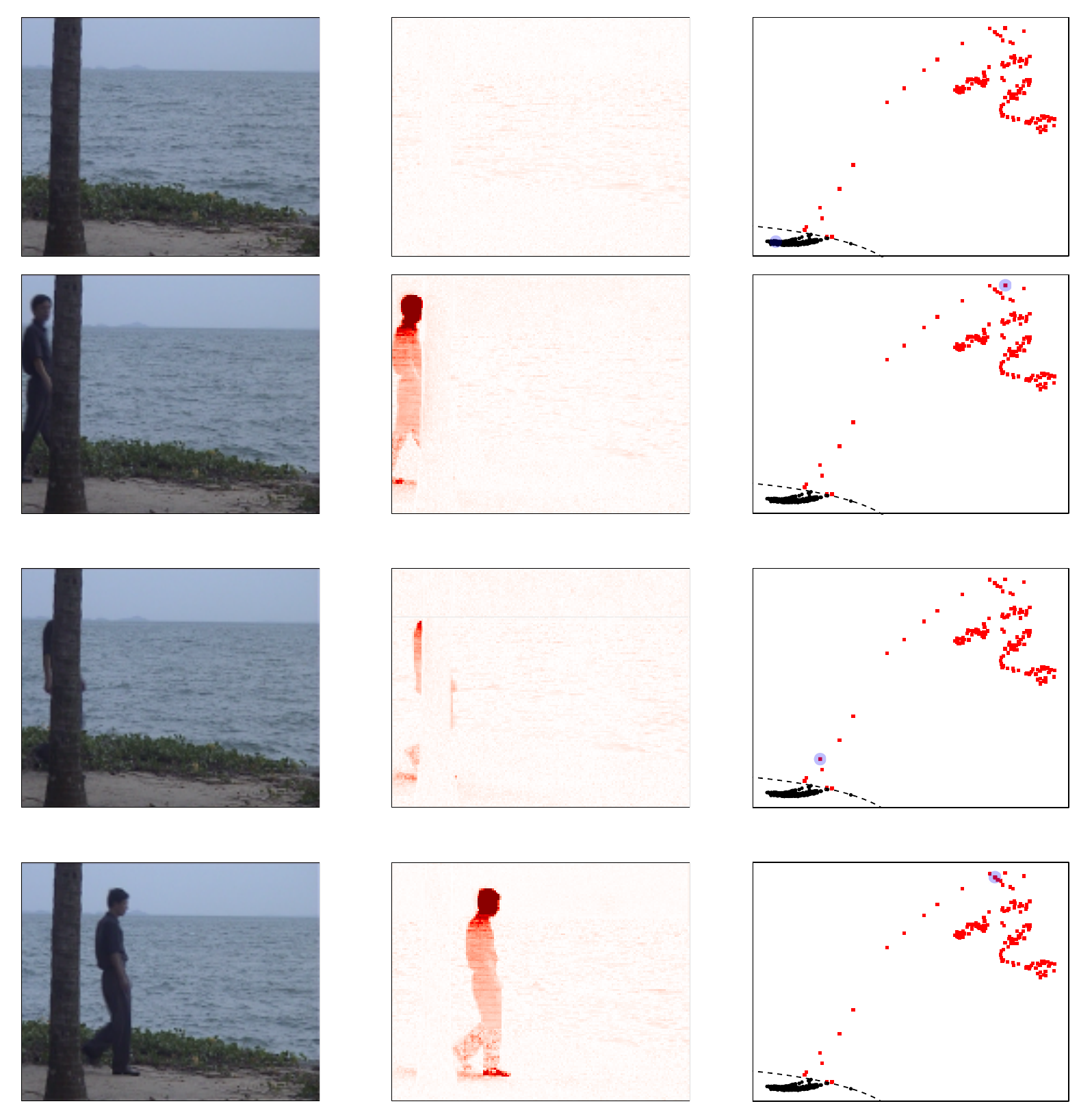}
\captionsetup{format=hang}
\caption{Left: Frames 100, 487, 491 and 500 from the video.
				 Middle: DO heatmaps of these frames.
				 Right: FOM with blue marker at the position of the
				 frame.}
\label{fig:videosum}
\end{figure}

In addition to the FOM we can draw DO heatmaps of the
individual frames.  
For frames 100, 487, 491 and 500, Figure~\ref{fig:videosum} 
shows the raw frame on the left, the DO heatmap in the middle 
and the FOM on the right, in which a blue circle marks the
position of the frame. 
In this figure we can follow the man's path in the FOM, 
while the DO heatmaps show exactly where the man is in 
those frames.    
We have created a video in which the raw frame, the DO 
heatmap and the FOM evolve alongside each other. 
It can be downloaded from 
{\it http://wis.kuleuven.be/stat/robust/publ}\;.

\section{Simulation Study}
We would also like to study the performance of the DO
when the data generating mechanism is known,
and compare it with the
AO measure proposed by \cite{Brys:RobICA} and studied
by \cite{Hubert:OutlierSkewed} and \cite{Hubert:MFOD}. 
For this we carried out an extensive simulation study,
covering univariate as well as multivariate and 
functional data.

\begin{figure}[!ht]
\centering
\includegraphics[width=0.95\textwidth]
  {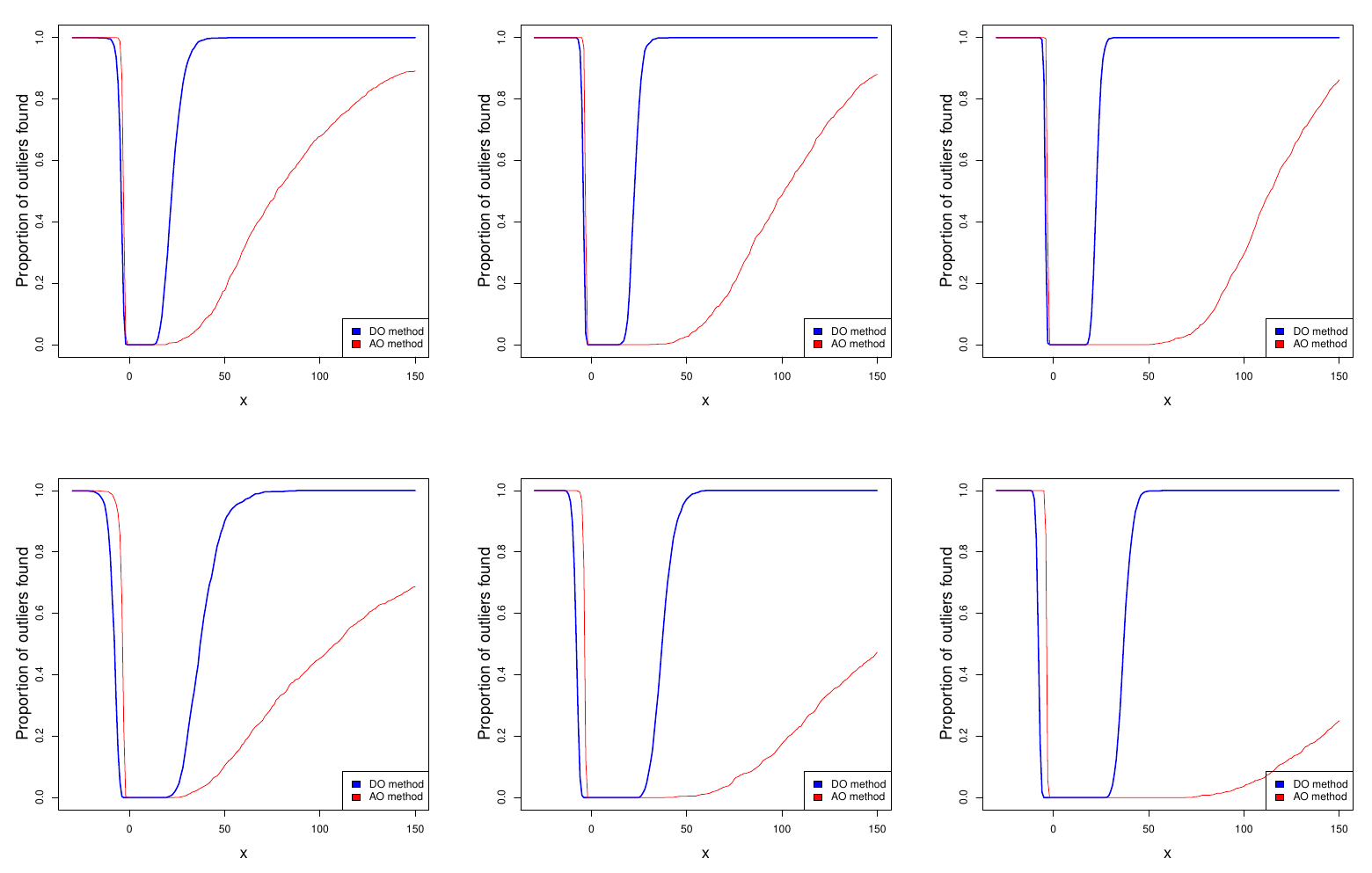}
\caption{Percentage of outliers found in univariate
  lognormal samples of size $n$=200 (left), $n$=500
	(middle), and $n$=1000 (right),	with 10\% (top) and 
	15\% (bottom) of outliers in $x$.}
		\label{fig:univariate_simulation}
\end{figure}

In the univariate case we generated 
$m=1000$ standard lognormal samples of size 
$n=\{200,500,1000\}$ with 10\% and 15\% of 
outliers at the position $x$, which may be negative.
Figure \ref{fig:univariate_simulation} shows 
the effect of the contamination at $x$ on our figure of 
merit, the percentage of outliers flagged (averaged over 
the $m$ replications).

In the direction of the short left tail of the lognormal
distribution we see that the adjusted outlyingness AO
flags about the same percentage of outliers as the DO.
But the AO is much slower in flagging outliers in the 
direction of the long right tail of the lognormal. 
This is due to the relatively high explosion bias of the 
scale used in the denominator of the AO for points to the 
right of the median.
The DO flags outliers to the right of the median much 
faster, due to its lower explosion bias.

\begin{figure}[!ht]
\centering
\includegraphics[width=0.95\textwidth]
  {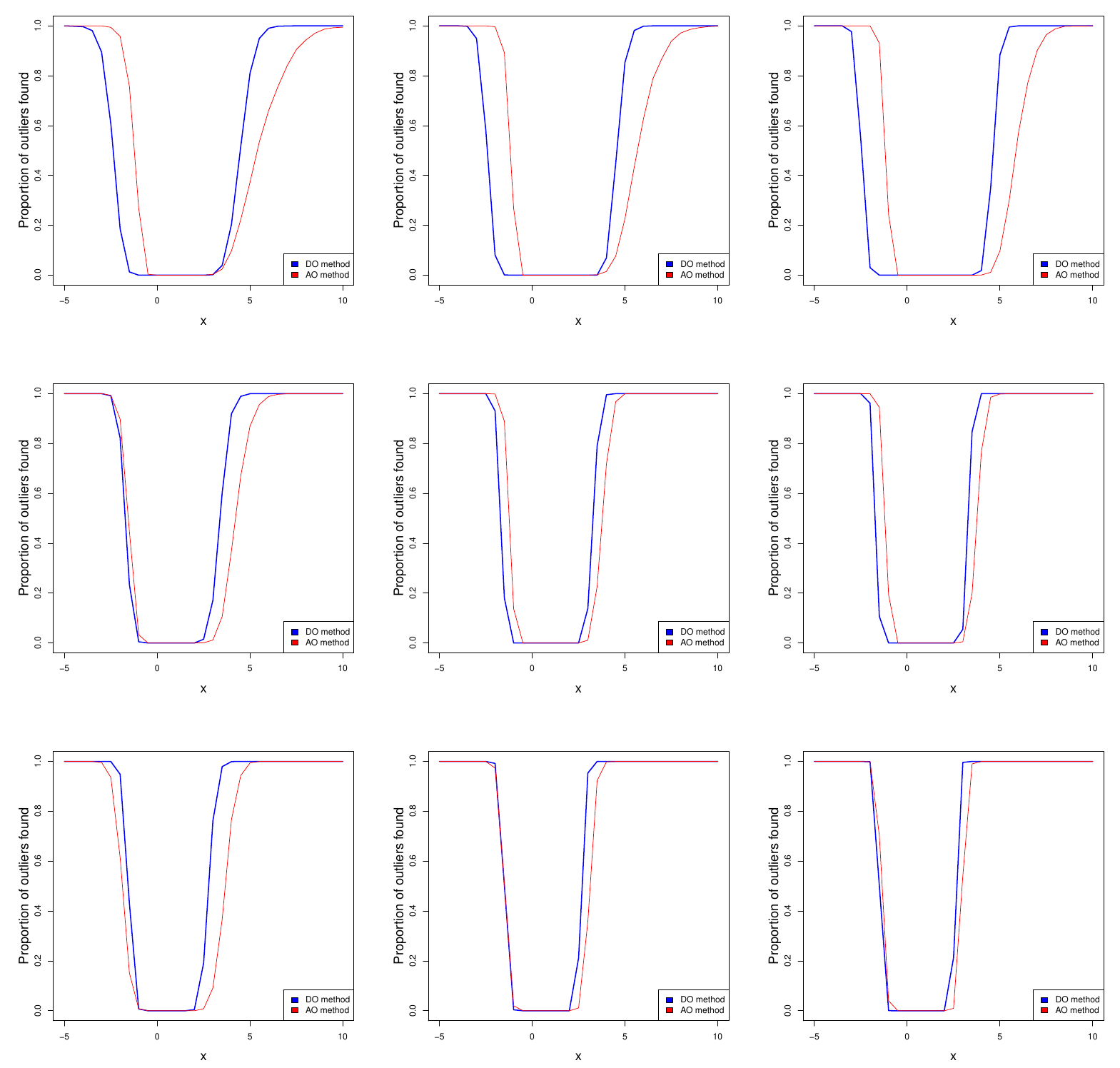}
\caption{Percentage of outliers found in multivariate
  skew normal samples of size $n$=200 (left), $n$=500
	(middle), and $n$=1000 (right), with 10\% of outliers
	around $\bx = (x,\ldots,x)^T$, 
	in dimensions $d$=2 (top), $d$=5 (middle),
	and $d$=10 (bottom).}
		\label{fig:multivariate_simulation}
\end{figure}

We have also extended the multivariate simulation of AO 
in \cite{Hubert:OutlierSkewed}.
Our simulation consists of $m=1000$ samples in dimensions 
$d=\{2,5,10\}$ and with sample sizes $n=\{200,500,1000\}$. 
The clean data were generated from the multivariate skew 
normal distribution \citep{Azz96} with density
$f(\by)=2\phi_d(\by) \Phi(\balpha^T \by)$
where $\Phi$ is the standard normal cdf, $\phi_p$ is the 
d-variate standard normal density, and $\balpha$
is a d-variate vector which regulates the shape. 
In our simulations $\balpha$ is a vector with entries equal 
to 10 or 4. 
For $d = 2$ we used $\balpha = (10, 4)^T$, for $d = 5$ we 
put $\balpha = (10, 10, 4, 4, 4)^T$, and for $d = 10$ we 
took $\balpha = (10, 10, 10, 10, 10, 4, 4, 4, 4, 4)^T$.
To this we added 10\% of contamination with a normal
distribution $N(\bx,I_d/20)$ around the point
$\bx=(x,...,x)^T$, where x is on the horizontal axis of
Figure \ref{fig:multivariate_simulation}.
In $d=2$ dimensions we see that AO flags the outliers a 
bit faster in the direction of the shortest tail, but slower
in the direction of the longest tail. 
The latter is similar to what we saw for univariate
data, due to the higher explosion bias of the
scale used (implicitly) in the AO.
When both the dimension $d$ and the sample size $n$ go up, 
the DO and AO methods give more similar results.
This is due to the fact that, in most directions, the
scales $s_a$ and $s_b$ of the projected data get closer 
to each other.
This is because the projections of the good data (i.e.
without the outliers) tend to become more gaussian as the
dimension $d$ and the sample size $n$ go up,
as shown by~\cite{DF84} for random directions
uniformly distributed on the unit sphere and under moment 
conditions on the data distribution.

\begin{figure}[!ht]
\centering
\includegraphics[width=0.5\textwidth]{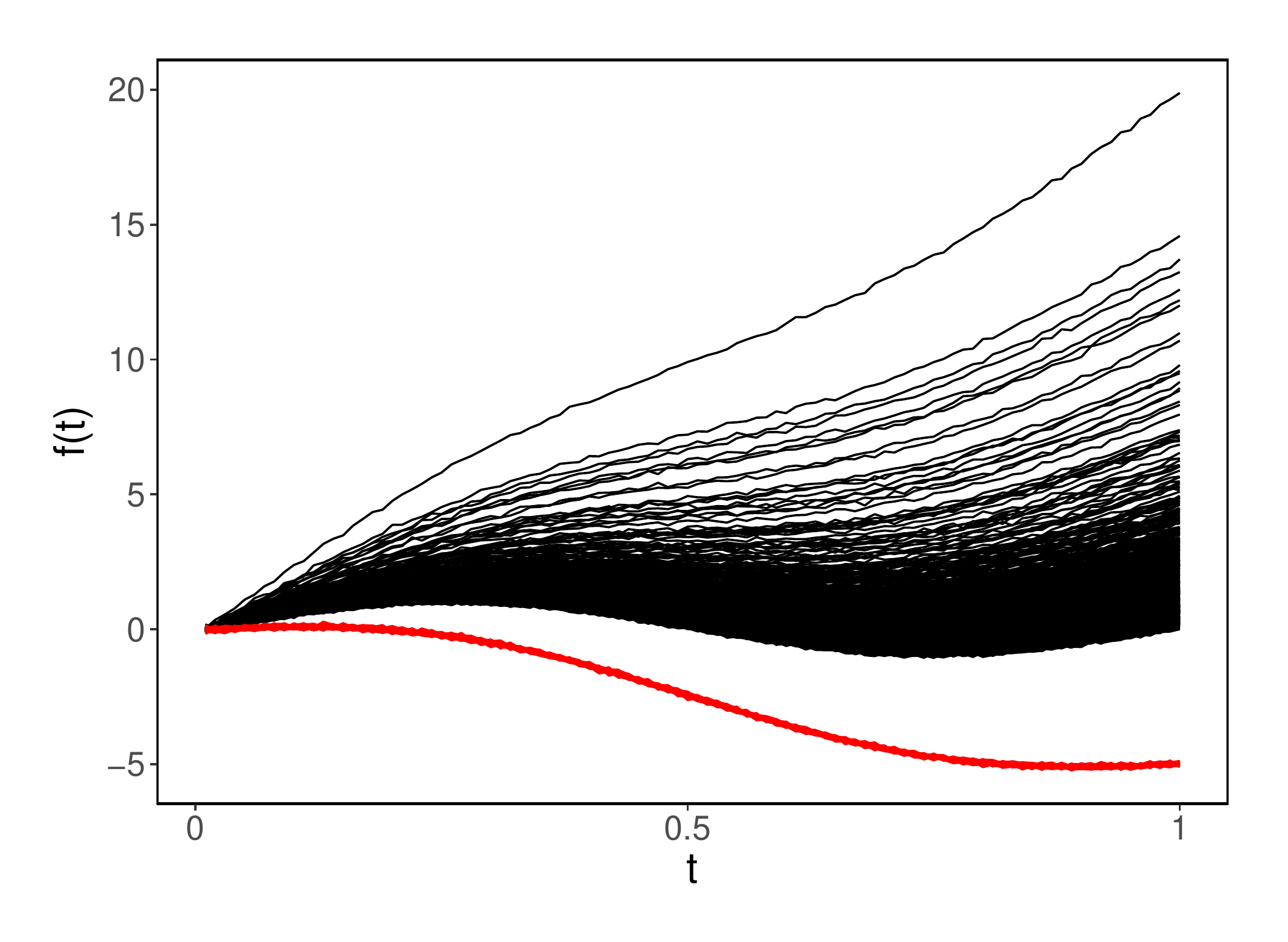}
\caption{$n=1000$ generated functions with 10\% 
         contamination.}
		\label{fig:func_example}
\end{figure}

We also carried out a simulation with functional data. 
We have generated $m=1000$ samples of $n=\{200,500,1000\}$ 
functions of the form
\begin{equation} \label{eq:sine}
  f_i(t)=\sin(2 \pi t)+t L_i+\varepsilon_i(t)
         \hspace{0.5cm} \mbox{for}
				 \hspace{0.5cm} 0 \leqslant t \leqslant 1
\end{equation}
where $\ln(L_i) \sim N(0,1)$ and 
$\varepsilon_i(t)\sim N\left(0,\left(\frac{1}{20}\right)^2\right)$. 
That is, the base function is the sine and we add different
 straight lines of which the slopes are generated by a 
 lognormal distribution.
We then replace 10\% of the functions by contaminated ones,
which are generated from~\eqref{eq:sine} but where $L_i$ 
is taken higher or lower than what one would expect under 
the lognormal model. 
Figure \ref{fig:func_example} shows such a generated data 
set of size $n=1000$, with outlying functions (with
negative $L_i$) in red.

\begin{figure}[!ht]
\centering
\includegraphics[width=0.95\textwidth]
  {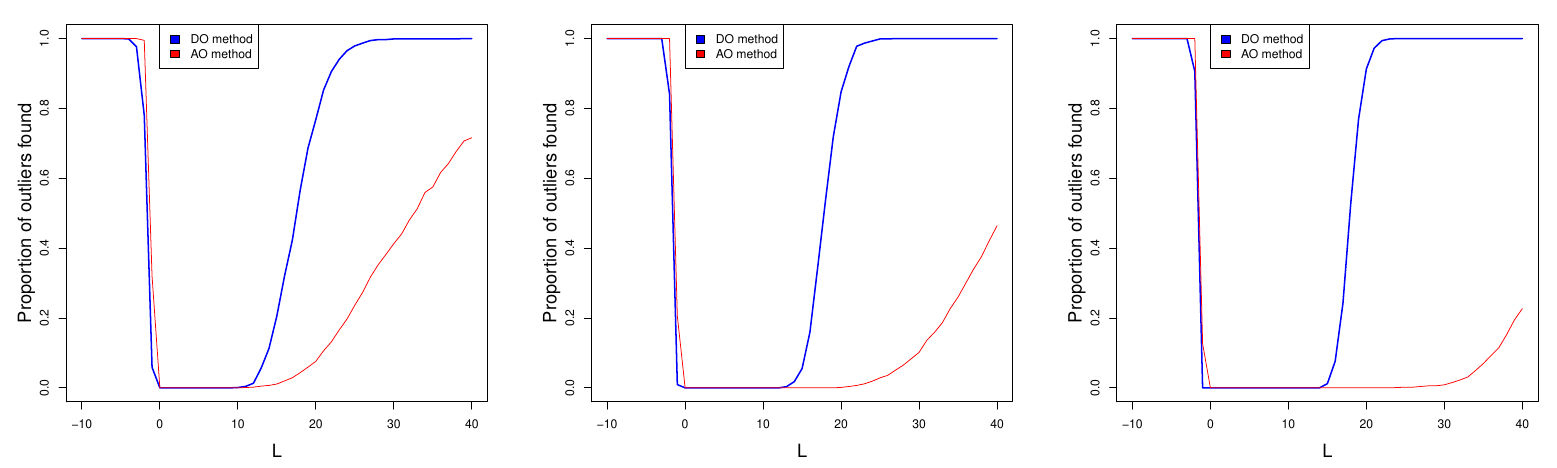}
\caption{Percentage of outliers found in functional samples
  of size $n$=200 (left), $n$=500 (middle), and $n$=1000
	(right), with 10\% of contaminated curves with slope $L$.}
		\label{fig:func_simulation}
\end{figure}

In the simulation we used a single slope $L$ for the
10\% of contaminated curves, and this $L$ is 
shown on the horizontal axis in 
Figure \ref{fig:func_simulation}.
When the outlying functions lie below the regular ones
(i.e. for negative $L$), we see that the DO and AO 
behave similarly.
On the other hand, when the outlying functions lie
above the regular ones (i.e. in the direction of the
long tail), the AO is much slower to flag them than DO.

These simulations together suggest that the DO outperforms
AO in directions where the uncontaminated data has
a longer tail, while performing similarly in the other
directions.

\begin{figure}[!ht]
\centering
\includegraphics[width=0.40\textwidth]{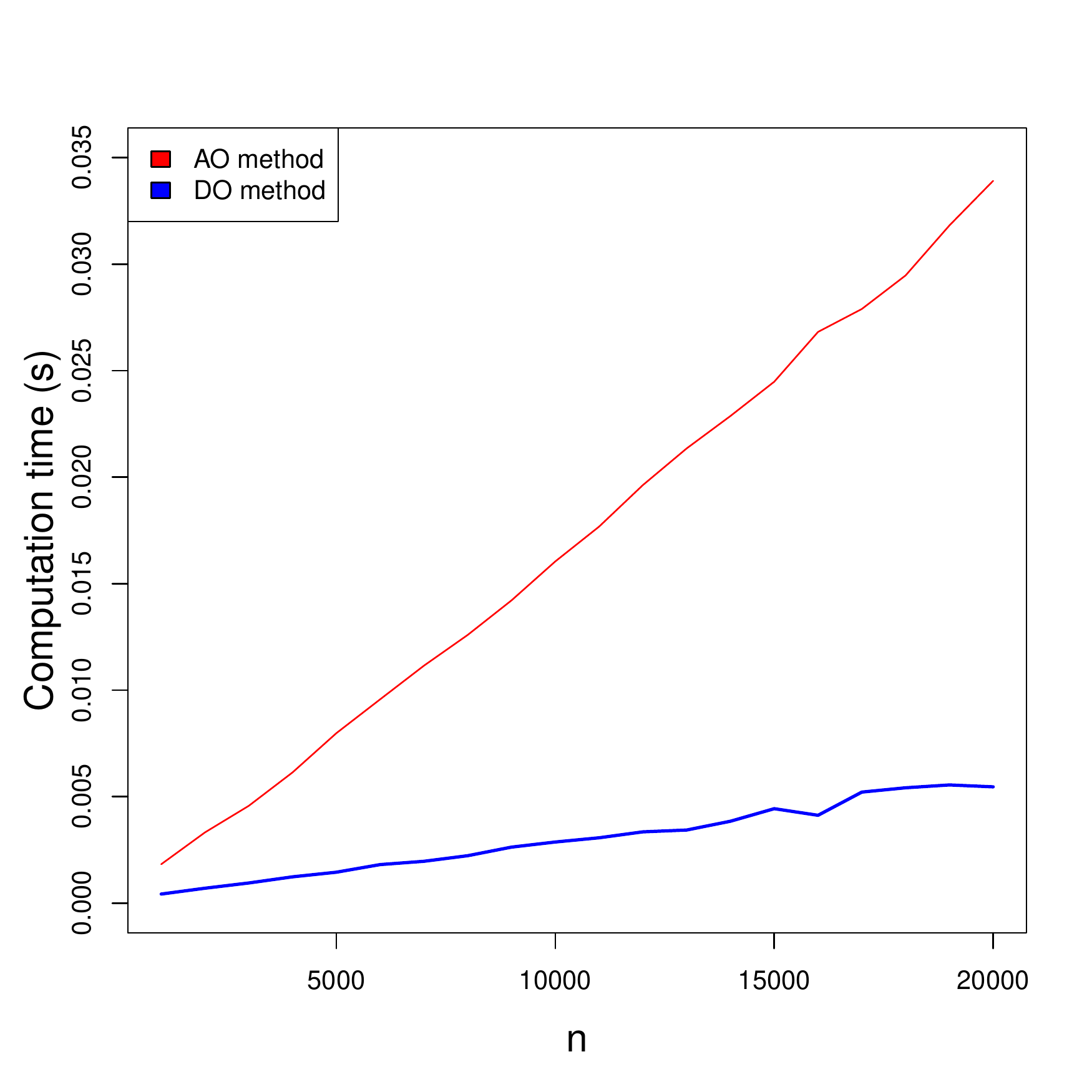}
\caption{Average computation time of DO and AO as a function
         of sample size.}
		\label{fig:comptime}
\end{figure}

Note that the DO requires only $\mathcal{O}(n)$ 
 computation time per
 direction, which is especially beneficial for functional 
 data with a large domain.
In particular, DO is much faster than AO which requires 
$\mathcal{O}(n \log(n))$ operations. 
Figure \ref{fig:comptime} shows the average computation 
time (in seconds) of both measures as a function of the 
sample size $n$, for $m=1000$ samples from the standard 
normal. 
The AO time is substantially above the DO time.

\section{Conclusion}
\label{sec:conc}
The notion of directional outlyingness (DO) is 
well-suited for skewed distributions. 
It has good robustness properties, and lends itself to
the analysis of univariate, multivariate, and functional 
data, in which both the domain and the function values 
can be multivariate.
Rough cutoffs for outlier detection are available.
The DO is also a building block of several graphical
tools like DO heatmaps, DO contours, and the functional
outlier map (FOM).
These proved useful when analyzing spectra, MRI images,
and surveillance video. 
In the MRI images we added gradients to the data in 
order to reflect shape/spatial information. 
In video data we could also add the 
numerical derivative in the time direction. 
In our example this would make the frames 6-dimensional, 
but the componentwise DO in~\eqref{eq:CDO} would remain 
fast to compute.

\section{Available software}
R-code for computing the 
DO and reproducing the examples is available
from our website 
{\it http://wis.kuleuven.be/stat/robust/software}\;.

\vskip0.5cm

\bigskip
\begin{center}
{\large\bf SUPPLEMENTARY MATERIAL}
\end{center}

\vskip0.2cm

\begin{proof}[\textbf{Proof of Lemma 1(i)}]
Let $\mu \in \R$ be fixed.
For the function $\rho_c$ we have that
\begin{align*}
\begin{array}{ll}
  t^2 \int_{\mu}^{\infty}{\rho_c\left(\frac{x-\mu}{t}\right)
	dF(x)}
  &=\;\; \int_{\mu}^{\infty}{\left\{\left(\frac{x-\mu}
	 {c}\right)^2 \one_{\left|\frac{x-\mu}{t}\right|\leq c}
	 +t^2 \one_{\left|\frac{x-\mu}{t}\right|>c} \right\}dF(x)}\\
  &=\;\; \int_{0}^{\infty}{\left\{\left(\frac{u}{c}\right)^2
	 \one_{0\leqslant u \leqslant ct}
	+t^2 \one_{ct<u}\right\} dF(\mu+u)}
\end{array}
\end{align*}
For all $u \geqslant 0$ it holds that
   $\left(\frac{u}{c}\right)^2
    \one_{0\leqslant u \leqslant ct} +t^2 \one_{ct<u}$ 
is nondecreasing in $t$, and even strictly increasing in $t$
at large enough $u$.
This proves (i) since $f(x) > 0$ in all $x$.
\end{proof}

\vskip0.2cm

\begin{proof}[\textbf{Proof of Lemma 1(ii)}]
Fix $\sigma>0$.
It follows from the Leibniz integral rule that
\begin{align*}
\begin{array}{ll}
  \frac{\partial}{\partial t}\left\{ \sigma^2 \int_{t}^{\infty}
    {\rho_c\left(\frac{x-t}{\sigma}\right) dF(x)}\right\}
  &=\; -\sigma \int_{t}^{\infty}{\rho_c'\left(
	    \frac{x-t}{\sigma}\right) dF(x)}
\end{array}
\end{align*}
because $\rho_c(0) = 0$.
Note now that 
\begin{equation*}
\begin{array}{lll}
  \rho_c'\left(\frac{x-t}{\sigma}\right) > 0 
  &\text{ for }& t \leqslant x<t+\sigma c\\
  \rho_c'\left(\frac{x-t}{\sigma}\right) = 0
	&\text{ for }& x>t+\sigma c \;\;.
\end{array}
\end{equation*}
This implies that
 $\frac{\partial}{\partial t}\left\{ \sigma^2 \int_{t}^{\infty}
  {\rho_c\left(\frac{x-t}{\sigma}\right) dF(x)}\right\} < 0$
for all $t$.
\end{proof}

\vskip0.2cm

\begin{proof}[\textbf{Proof of Proposition 1}]
Let $0<\varepsilon<0.25$ be fixed and let 
$F_{\varepsilon,H}$ be a minimizing distribution, 
i.e.
\begin{align*}
 \inf_{F_{\varepsilon,G} \in \mathcal{F}_{\varepsilon,G}}
 (\sam(F_{\varepsilon,G}))=\sam(F_{\varepsilon,H})
\end{align*}
with $F_{\varepsilon,G}=(1-\varepsilon)F+\varepsilon \;G$\;.
Inserting the contaminated distribution $F_{\varepsilon,H}$ 
into $\sam(F_{\varepsilon,H})$ 
yields the scale
\begin{equation}\label{eq:samFeH}
\begin{aligned}
  \frac{\soa^2(F_{\varepsilon,H})}{\alpha} 
	\left\{(1-\varepsilon) \; 
	\int_{\med (F_{\varepsilon,H})}^{\infty}
	{\rho_c \left(\frac{x-\med(F_{\varepsilon,H})}
	{\soa (F_{\varepsilon,H})}\right) dF(x)} \right.\\
  \left. +\;\varepsilon 
	\int_{\med (F_{\varepsilon,H})}^{\infty}
	{\rho_c \left(\frac{x-\med(F_{\varepsilon,H})}
	{\soa (F_{\varepsilon,H})}\right) dH(x)}\right\} \;\;.
\end{aligned}
\end{equation}
For simplicity, put
\begin{equation}\label{eq:simplif}
\begin{aligned}
 W_1(F_{\varepsilon,H}) &=
   \int_{\med (F_{\varepsilon,H})}^{\infty}
	 {\rho_c \left(\frac{x-\med(F_{\varepsilon,H})}
	 {\soa (F_{\varepsilon,H})}\right) dF(x)}\\
 W_2(F_{\varepsilon,H}) &=
   \int_{\med (F_{\varepsilon,H})}^{\infty}
	 {\rho_c \left(\frac{x-\med(F_{\varepsilon,H})}
	 {\soa (F_{\varepsilon,H})}\right) dH(x)}\;\;. 
\end{aligned}
\end{equation}
We then have the contaminated scale 
\begin{equation}\label{eq:shorter}
\frac{\soa^2(F_{\varepsilon,H})}{\alpha} \left\{ 
  (1-\varepsilon) \; W_1(F_{\varepsilon,H})
	+\varepsilon W_2(F_{\varepsilon,H})\right\}.
\end{equation}
Denote by $Q_{2,\varepsilon}=F_{\varepsilon,H}^{-1}(0.5)$
and $Q_{3,\varepsilon}=F_{\varepsilon,H}^{-1}(0.75)$ 
the median and the third quartile of the contaminated
distribution.

For the distribution $H$ it has to hold that 
$H(Q_{3,\varepsilon})=1$ and 
$\lim_{x \to Q_{2,\varepsilon}^{-}}{H(x)}=0$. 
This can be seen as follows.
Suppose $H(\infty)-H(Q_{3,\varepsilon})=p \in (0,1]$.
Then consider $F_{\varepsilon,H^*}$\; where
$$H^*(x)=
\left\{
\begin{array}{ll}
H(x)+p\Delta(Q_{2,\varepsilon}) & \text{ for } x \in
   (-\infty,Q_{3,\varepsilon}] \\
    1& \text{ else }\\
\end{array}
\right.$$ 
and denote by $Q_{2,\varepsilon}^*$ and $Q_{3,\varepsilon}^*$ 
the median and third quartile of $F_{\varepsilon,H^*}$\;.
Note that $Q_{2,\varepsilon}=Q_{2,\varepsilon}^*$ and 
$Q_{3,\varepsilon}>Q_{3,\varepsilon}^*$.
Therefore, we have 
$\soa(F_{\varepsilon,H}) > \soa(F_{\varepsilon,H^*})$. 
It then follows from Lemma 1(i) that 
$\soa(F_{\varepsilon,H})^{2} (1-\varepsilon) 
  W_1(F_{\varepsilon,H})>
  \soa(F_{\varepsilon,H^*})^{2}
  (1-\varepsilon) W_1(F_{\varepsilon,H^*})$ 
and 
 $W_2(F_{\varepsilon,H})>W_2(F_{\varepsilon,H^*})=0$. 
Therefore 
 $\soa(F_{\varepsilon,H})^{2} \varepsilon 
  W_2(F_{\varepsilon,H}) > \soa(F_{\varepsilon,H^*})^{2} 
	\varepsilon W_2(F_{\varepsilon,H^*})$.
It now follows that 
$\sam (F_{\varepsilon,H^*})<\sam (F_{\varepsilon,H})$, 
which is a contradiction since $H$ minimizes $\sam$. 
Therefore, $H(\infty)-H(Q_{3,\varepsilon})=0$. 
A similar argument can be made to show that 
  $\lim_{x \to Q_{2,\varepsilon}^{-}}{H(x)}=0$. 
It follows that $H(Q_{3,\varepsilon})=1$ and 
  $H(x)=0$ for all $x < Q_{2,\varepsilon}$,
so all the mass of $H$ is inside
  $[Q_{2,\varepsilon}\,,\,Q_{3,\varepsilon}]$\;.
	
We can now argue that $H$ must have all its mass in 
$Q_{2,\varepsilon}$.
Note that if $H(Q_{3,\varepsilon})=1$ and 
 $\lim_{x \to Q_{2,\varepsilon}^{-}}{H(x)}=0$ 
we have 
 $Q_{2,\varepsilon}=F^{-1}\left(\frac{1}
 {2(1-\varepsilon)}\right)$ 
and 
  $Q_{3,\varepsilon} \in \left[F^{-1}\left(
	 \frac{3-4\varepsilon}{4(1-\varepsilon)}\right),
	 F^{-1}\left(\frac{3}{4(1-\varepsilon)}\right)\right]$, 
depending on $\lim_{x \to Q_{3,\varepsilon}^{-}}{H(x)}$.
Given that $Q_{2,\varepsilon}$ is fixed, we can minimize 
$W_1(F_{\varepsilon,H})$ by minimizing $Q_{3,\varepsilon}$. 
Now $Q_{3,\varepsilon}$ is minimal for 
  $H=\Delta \left(F^{-1}\left(\frac{1}
	 {2(1-\varepsilon)}\right)\right)$ 
as this yields
	$Q_{3,\varepsilon} = F^{-1}\left(
	 \frac{3-4\varepsilon}{4(1-\varepsilon)}\right)$. 
Note that this choice of $H$ to minimize $Q_{3,\varepsilon}$ 
is not unique as any $H$ which makes 
$\lim_{x \to Q_{3,\varepsilon}^{-}}{H(x)}=1$ does the job.
Note finally that $W_2(F_{\varepsilon,H})$ 
is also minimal for 
  $H=\Delta \left(F^{-1}\left(
	 \frac{1}{2(1-\varepsilon)}\right)\right)$ 
as $\rho_c(t)$ is nondecreasing in $|t|$, and this choice 
of $H$ yields $W_2(F_{\varepsilon,H})=0$.

We now know that $H=\Delta \left(F^{-1}\left(
 \frac{1}{2(1-\varepsilon)}\right)\right)$ 
minimizes $\sam(F_{\varepsilon,H})$.
Furthermore, we have
  $Q_{2,\varepsilon}=F^{-1}\left(\frac{1}{2(1-\varepsilon)}
	\right)=B^+(\varepsilon,\med,F)$ and
	$Q_{3,\varepsilon}= F^{-1}\left(\frac{3-4\varepsilon}
	{4(1-\varepsilon)}\right)$. 
Therefore the implosion bias of $\sam$ is
\begin{align*}
\begin{array}{ll}
  B^-(\varepsilon,\sam,F)^2
	&=\;\frac{B^-(\varepsilon,\soa,F)^2}{\alpha}
	\left\{ (1-\varepsilon) 
	\int\displaylimits_{B^+(\varepsilon,\med,F)}^{\infty}
	{\rho_c\left(\frac{x-B^+(\varepsilon,\med,F)}
	{B^-(\varepsilon,\soa,F)}\right)dF(x)} \right\}
\end{array}
\end{align*}
where
\begin{equation*}
\begin{array}{rl}
  B^+(\varepsilon,\med,F)
    &= \;\; F^{-1}\left(\frac{1}{2(1-\varepsilon)}\right)\\
  B^-(\varepsilon,\soa,F)
	  &= \;\; \left(F^{-1}\left(\frac{3-4\varepsilon}
		 {4(1-\varepsilon)}\right)
     -F^{-1}\left(\frac{1}{2(1-\varepsilon)}\right)\right)
     /\Phi^{-1}(0.75)\;\;.
\end{array}
\end{equation*}
\end{proof}

\vskip0.2cm

\begin{proof}[\textbf{Proof of Proposition 2}]
Let $0<\varepsilon<0.25$ be fixed and let 
$F_{\varepsilon,H}$ be a maximizing distribution, 
i.e.
\begin{align*}
 \sup_{F_{\varepsilon,G} \in \mathcal{F}_{\varepsilon,G}}
 (\sam(F_{\varepsilon,G}))=\sam(F_{\varepsilon,H})
\end{align*}
with $F_{\varepsilon,G}=(1-\varepsilon)F+\varepsilon \;G$\;.
Inserting the contaminated distribution $F_{\varepsilon,H}$ 
into $\sam(F_{\varepsilon,H})$ yields the 
scale~\eqref{eq:samFeH}, which can be rewritten as 
in~\eqref{eq:simplif} and~\eqref{eq:shorter}.

For the distribution $H$ it has to hold that 
 $H(Q_{3,\varepsilon})=
  \lim_{x \to Q_{2,\varepsilon}^{-}}{H(x)}$. 
This can be seen as follows. Suppose 
 $H(Q_{3,\varepsilon})-\displaystyle
  \lim_{x \to Q_{2,\varepsilon}^{-}}{H(x)}=p \in (0,1]$. 
Now put
$e=B^{+}(\varepsilon,\med,F) + cB^{+}(\varepsilon,\soa,F)$
and consider the distribution $F_{\varepsilon,H^*}$ where
$$H^*(x)=
\left\{
\begin{array}{ll}
  H(x) 
	& \text{ for } x \in (-\infty,Q_{2,\varepsilon})\\
  \displaystyle \lim_{x \to Q_{2,\varepsilon}^{-}}{H(x)}
	& \text{ for } x \in 
  [Q_{2,\varepsilon},Q_{3,\varepsilon}]\\
  H(x) - p + p\Delta(e)
	& \text{ for } x \in (Q_{3,e},\infty)\\
\end{array}
\right.$$
and denote by $Q_{2,\varepsilon}^*$ and $Q_{3,\varepsilon}^*$ 
the median and third quartile of $F_{\varepsilon,H^*}$\;.
Note that $Q_{2,\varepsilon}=Q_{2,\varepsilon}^*$ and 
$Q_{3,\varepsilon}<Q_{3,\varepsilon}^*$.
Therefore $\soa(F_{\varepsilon,H})<\soa(F_{\varepsilon,H^*})$ 
and thus
  $\soa(F_{\varepsilon,H})^{2} (1-\varepsilon) 
 	 W_1(F_{\varepsilon,H}) < \soa(F_{\varepsilon,H^*})^{2}
	 (1-\varepsilon) W_1(F_{\varepsilon,H^*})$ 
because of Lemma 1(i). 
Furthermore, 
  $W_2(F_{\varepsilon,H}) < W_2(F_{\varepsilon,H^*})$ 
because $\rho_c(t)$ is nondecreasing in $|t|$, thus 
  $\soa(F_{\varepsilon,H})^{2} \varepsilon 
	 W_2(F_{\varepsilon,H}) <
	 \soa(F_{\varepsilon,H^*})^{2} 
	 \varepsilon W_2(F_{\varepsilon,H^*})$.
It now follows that 
 $\sam(F_{\varepsilon,H^*}) > \sam(F_{\varepsilon,H})$, 
which is a contradiction since $H$ maximizes $\sam$.
Therefore $H(Q_{3,\varepsilon})=
  \lim_{x \to Q_{2,\varepsilon}^{-}}{H(x)}$\,,
so $H$ has no mass inside
  $[Q_{2,\varepsilon}\,,\,Q_{3,\varepsilon}]$\;.

Without loss of generality we can thus assume that $H$ is
of the form $H=d\Delta(e_1) + (1-d)\Delta(e_2)$ with
 $e_1=B^{-}(\varepsilon,\med,F) - 
      cB^{+}(\varepsilon,\soa,F)$ and 
 $e_2=B^{+}(\varepsilon,\med,F) +
      cB^{+}(\varepsilon,\soa,F)$ 
where $d \in [0,1]$. 
This choice of $e_1$ and $e_2$ is not unique but it 
maximizes $\sam(F_{\varepsilon,H})$ because $\rho_c(t)$ 
is nondecreasing in $|t|$.
Inserting the distribution $F_d := F_{\varepsilon,H}$ yields
\begin{align}\label{eq:sam2d}
  \sam^2(F_d)=
	  \frac{\soa^2(F_d)}{\alpha}\left\{
		(1-\varepsilon) \int_{Q_{2,d}}^{\infty}
		{\rho_c \left(\frac{x-Q_{2,d}}
		{\soa(F_d)}\right) dF(x)}
		+\varepsilon(1-d) \right\}
\end{align}
where 
  $Q_{2,d}=F^{-1}\left(
	  \frac{1-2d\varepsilon}{2(1-\varepsilon)}\right)$,
	$Q_{3,d}=F^{-1}\left(
	  \frac{3-4d\varepsilon}{4(1-\varepsilon)}\right)$ and
	$\soa (F_d)=\left(Q_{3,d}-
	  Q_{2,d}\right)/\Phi^{-1}(0.75)$\;.
Note that this expression depends on $d$
but no longer on $e_1$ and $e_2$.
We will show that this expression is maximized for $d=0$.

First we show that $\soa(F_d)$ is maximized for 
$d=0$. Let 
\begin{equation}\label{eq:g}
  g(d) := \soa(F_d)=(Q_{3,d}-Q_{2,d})/\Phi^{-1}(0.75)
\end{equation}
for any $\;d\in [0,1]\;$. Note that 
  $\xi=\frac{3-4\varepsilon d}{4(1-\varepsilon)}-
	\frac{2-4\varepsilon d}{4(1-\varepsilon)}
	=\frac{1}{4(1-\varepsilon)}$ 
does not depend on $d$. Therefore, we can write
  $g(d)=(F^{-1}(v+\xi)-F^{-1}(v))/\Phi^{-1}(0.75)$ 
where $v=\frac{2-4\varepsilon d}{4(1-\varepsilon)}$ is 
a strictly decreasing function of $d$.
Note that we can write 
  $g(d)=(\Phi^{-1}(0.75))^{-1}
	      \int_{v}^{v+\xi}{\frac{d u}{f(F^{-1}(u))}}$. 
The density $f$ is symmetric about some $m$, and by 
affine equivariance we can assume $m=0$ w.l.o.g.
Since $f$ is unimodal with $f(x)>0$ for all $x$, the 
function $u\to \frac{1}{f(F^{-1}(u))}$ is  
strictly decreasing up to its minimum (corresponding to 
the mode of $f$) and then strictly increasing.
Therefore, $g(d)$ is maximal for $v$ as large as possible, 
i.e. for $d=0$. In that case, we have 
$v=Q_{2,o}=\frac{2}{4(1-\varepsilon)}>0.5\;$.

Next, we maximize
\begin{equation} \label{eq:h} 
h(\sigma,d) := \frac{\sigma^2}{\alpha}\left\{
		(1-\varepsilon) \int_{Q_{2,d}}^{\infty}
		{\rho_c \left(\frac{x-Q_{2,d}}
		{\sigma}\right) dF(x)}
		+\varepsilon(1-d) \right\}
\end{equation}
for any fixed $\sigma>0$\;.
This is equivalent to maximizing
\begin{equation}\label{eq:objective}
  \int_{q}^{\infty}{\rho_c\left(\frac{x-q}{\sigma}\right)dF(x)
  +\frac{\varepsilon}{1-\varepsilon}(1-d)}
\end{equation}
where $q$ is such that  
$F(q)\in \left[\frac{1-2\varepsilon}{2(1-\varepsilon)},
  \frac{1}{2(1-\varepsilon)}\right]=\frac{1}{2}\pm
	\frac{\varepsilon}{1-\varepsilon}\;$. 
Note that 
$\frac{\varepsilon}{1-\varepsilon}(1-d)
 =F(q)+\frac{1-2\varepsilon}{2(1-\varepsilon)}$, 
where the second term doesn't depend on $q$. 
Maximizing~\eqref{eq:objective} with respect to $q$ is 
therefore equivalent to maximizing 
  $\int_{q}^{q+c \sigma}{\left(\frac{x-q}{c\sigma}\right)^2
	 dF(x)}-\int_{q}^{q+c \sigma}{dF(x)}\;$. 
Note that this is equal to 
  $\int_{0}^{c\sigma}{\left(\frac{x}{c\sigma}\right)^2dF(q+x)}
	-\int_{0}^{c\sigma}{dF(q+x)}
	=\int_{0}^{c\sigma}{\frac{x^2-\sigma^2c^2}{\sigma^2 c^2}
	f(q+x) d(x)}$. 
For all $x$ in $[0,c\sigma]$ it holds that 
  $\frac{x^2-\sigma^2c^2}{\sigma^2 c^2} \leq 0$,
hence the latter integral is maximized by minimizing 
$f(q+x)$ for all $x \in [0,c\sigma]$. 
For this $q$ must take on its highest possible value
$q=F^{-1}\left(\frac{1}{2}+\frac{\varepsilon}
 {1-\varepsilon}\right)\;$, 
because we then have $f(q+x) \leq f(q_2+x)$
for all $x$ in $[0,c\sigma]$ and all $q_2$ in 
$\left[F^{-1}\left(\frac{1}{2}-\frac{\varepsilon}
 {1-\varepsilon}\right),
 F^{-1}\left(\frac{1}{2}+\frac{\varepsilon}
 {1-\varepsilon}\right)\right]\;$.
Therefore, \eqref{eq:h} is maximized for $d=0$.

We now know that \eqref{eq:g} and \eqref{eq:h} satisfy
$\max_d g(d) = g(0)$ and $\max_d h(\sigma,d) = h(\sigma,0)$
for all $\sigma > 0$. 
By Lemma 1(i), $h(\sigma,0)$ is increasing in $\sigma$.
Combining these results yields
\begin{align*}
\begin{array}{ll}
  \max_{d}\;\sam^2(F_d)
	  &=\;\max_{d}\;h(g(d),d)\\
		&\leqslant\; \max_{d_1} \max_{d_2}\;h(g(d_1),d_2)\\
		&=\;\max_{d_1} \;h(g(d_1),0)\\
		&=\;h(g(0),0)\\
\end{array}
\end{align*}
so $\sam^2(F_d)$ is maximized for $d=0$.
Therefore, $\sam^2(F_{\varepsilon,H})$ is 
maximized for $H=\Delta(e_2)$, hence
 $Q_{2,\varepsilon}=F^{-1}\left(
  \frac{1}{2(1-\varepsilon)}\right)=
	B^+(\varepsilon,\med,F)$ and
	$Q_{3,\varepsilon}= F^{-1}\left(
	\frac{3}{4(1-\varepsilon)}\right)$. 
The explosion bias is thus
\begin{equation*}
\begin{array}{ll}
  B^+(\varepsilon,\sam,F)^2&=\;
	\frac{\soa^2}{\alpha}
	\left\{(1-\varepsilon) 
	\int\displaylimits_{B^+(\varepsilon,\med,F)}^{\infty}
	{\rho_c\left(\frac{x-B^+(\varepsilon,\med,F)}
	{\soa}\right)dF(x)}
	+\varepsilon \right\}\\
\end{array}
\end{equation*}
where
\begin{equation*}
\begin{array}{rl}
  \soa
	  &=\;\; \left\{F^{-1}\left(\frac{3}{4(1-\varepsilon)}\right)
		  -F^{-1}\left(\frac{1}{2(1-\varepsilon)}\right)\right\}/
			\Phi^{-1}(0.75)\;\;.
\end{array}
\end{equation*}
\end{proof}

\vskip0.2cm

\begin{proof}[\textbf{Proof of Proposition 3}]
Plugging the contaminated distribution 
$F_{\varepsilon,z}=(1-\varepsilon)F+\varepsilon \Delta z$ 
into the functional form~\eqref{eq:funcsasb} of $\sam$
yields
\begin{align*}
  \sam^2(F_{\varepsilon,z})=
  \frac{\soa^2(F_{\varepsilon,z})}{\alpha}
	\int_{\med (F_{\varepsilon,z})}^{\infty}
	{\rho_c \left(\frac{x-\med(F_{\varepsilon,z})}
	{\soa (F_{\varepsilon,z})}\right) dF_{\varepsilon,z}(x)}\;\;.
\end{align*}
We take the derivative with respect to $\varepsilon$ and 
evaluate it in $\varepsilon=0$.
Note that $\rho_c(t)$ is not differentiable at $t=c$ and 
$t=-c$, but as these two points form a set of measure zero 
this does not affect the integral containing $\rho_c'(t)$.
We also use that $\rho_c(0)=0$ and 
$\mbox{IF}(z,\sam^2,F)=2\sam(F)\mbox{IF}(z,\sam,F)$\;
yielding the desired expression.
For $F=\Phi$ we have
 $\mbox{IF}\left(z,\soa,\Phi\right)=
 \left(\one_{\{[0,\infty)\}}(z)\;
 \sign(z-\Phi^{-1}\left(\frac{3}{4}\right))
 +\mbox{IF} \left(z,\med,\Phi\right)
 \left\{\phi (0)-2\phi\left(\Phi^{-1}\left(
 \frac{3}{4}\right)\right)\right\}\right)/
 \left(2 \; \phi \left(\Phi^{-1}\left(
 \frac{3}{4}\right)\right)\right)\;.$\\
\end{proof}

\vskip0.2cm

\begin{proof}[\textbf{Proof of Proposition 4}]
We show the proof for $x>\med(F)$, the other case being 
analogous. Plugging the contaminated distribution 
$F_{\varepsilon,z}=(1-\varepsilon)F+\varepsilon \Delta z$ 
into $\DO(x,F)=(x-\med(F))/\sam(F)$ yields
\begin{align*}
\mbox{IF}(z,\DO(x),F) 
  &=\frac{\partial}{\partial \varepsilon}
	  (\DO(x,F_{\varepsilon,z}))\bigg|_{\varepsilon=0}\\
  &=\frac{1}{\sam^2(F_{\varepsilon,z})}
	  \left(-\frac{\partial}{\partial \varepsilon}\left(
		\med(F_{\varepsilon,z})\right) 
		\sam(F_{\varepsilon ,z})-
		\frac{\partial}{\partial \varepsilon}
		\left(\sam(F_{\varepsilon,z})\right)
		(x-\med(F_{\varepsilon,z}))\right)
		\bigg|_{\varepsilon=0}\\
  &=\frac{1}{\sam^2(F)}\left(-\mbox{IF}(z,\med,F) \;
	  \sam(F)-\mbox{IF}(z,\sam,F) \; (x-\med(F))\right)\;\;.
\end{align*}
\end{proof}

\end{document}